\documentclass[fleqn,usenatbib]{mnras}

\usepackage{newtxtext,newtxmath}

\usepackage[T1]{fontenc}

\DeclareRobustCommand{\VAN}[3]{#2}
\let\VANthebibliography\thebibliography
\def\thebibliography{\DeclareRobustCommand{\VAN}[3]{##3}\VANthebibliography}

\usepackage{graphicx}	
\usepackage{subfig}
\usepackage{amsmath}	
\usepackage{multicol}   
\usepackage{siunitx}
\usepackage{pdflscape}	

\title[Super-Slow Rotating Asteroids]{Discovery of Super-Slow Rotating Asteroids with ATLAS and ZTF photometry}

\author[N. Erasmus et al.]{
N. Erasmus,$^{1}$\thanks{E-mail: nerasmus@saao.ac.za}
D. Kramer,$^{2,3}$
A. McNeill,$^{3}$
D. E. Trilling,$^{3,1}$
P. Janse van Rensburg,$^{1,4}$
\newauthor{}
G. T. van Belle,$^{5}$
J. L. Tonry,$^{6}$
L. Denneau,$^{6}$
A. Heinze,$^{6}$
and H. J. Weiland$^{6}$ 
\\
$^{1}$South African Astronomical Observatory, Cape Town, 7925, South Africa\\
$^{2}$School of Informatics, Computing, and Cyber Systems, Northern Arizona University, Flagstaff, AZ 86011, USA\\
$^{3}$Department of Astronomy and Planetary Science, Northern Arizona University, Flagstaff, AZ 86011, USA\\
$^{4}$Department of Astronomy, University of Cape Town, Rondebosch, 7701, South Africa\\
$^{5}$Lowell Observatory, Flagstaff, AZ 86001, USA\\
$^{6}$Institute for Astronomy, University of Hawaii, Honolulu, HI 9682, USA.
}

\date{Accepted XXX. Received YYY; in original form ZZZ}

\pubyear{2020}

\begin{document}
\label{firstpage}
\pagerange{\pageref{firstpage}--\pageref{lastpage}}
\maketitle

\begin{abstract}
We present here 
the discovery of a new class of super-slow rotating asteroids (P$_{rot}\gtrsim$1000 hours) in data extracted from the Asteroid Terrestrial-impact Last Alert System (ATLAS) and Zwicky Transient Facility (ZTF) 
all-sky surveys.
Of the 39 rotation periods we report here, 32 have periods longer than any previously reported unambiguous rotation periods currently in the Asteroid Light Curve Database. In our sample, 7 objects have a rotation period $>$ 4000~hours and the longest period we report here is 4812 hours ($\sim$ 200~days). We do not observe any correlation between taxonomy, albedo, or orbital properties with super-slow rotating status. 
The most plausible mechanism for the creation of these very slow rotators is if their rotations were slowed by YORP spin-down. 
Super-slow rotating asteroids may be common, with at least 0.4\% of the main-belt asteroid population with a size range between 2 and 20~km in diameter rotating with periods longer than 1000~hours.
\end{abstract}

\begin{keywords}
minor planets, asteroids: general
\end{keywords}



\section{Introduction}

There are over 1 million known asteroids as of this writing\footnote{\url{https://www.minorplanetcenter.net/}}, but for the vast majority of these nothing is known beyond their orbital elements. Asteroids can act as tracers of the formation and evolution of the Solar System, and therefore understanding their physical properties helps improve our understanding of the processes that have shaped our planetary system from its formation to today. For example, the lightcurve of an individual asteroid can be used to constrain that body's intrinsic strength for certain combinations of lightcurve amplitude and period. In addition, the lightcurve properties of a large sample of asteroids can be used to calculate the mean shape of that population (\citealt{McNeill2016}, \citealt{McNeill2019}) and, in some cases, e.g., Jupiter Trojans, reveal the history of that population (\citealt{McNeill2021}). Thus, both individual and ensemble asteroid lightcurve properties are valuable in understanding how our Solar System formed and evolved.

Observing asteroids individually is time-consuming, especially when measuring properties that require large time investments, such as determining rotation periods through lightcurves. For this reason, the number of asteroids with very long rotation periods has been essentially unknown. Of the more than 1 million known asteroids there are currently only 8021 asteroids with reported rotation periods in the Asteroid Light Curve Database\footnote{\url{http://alcdef.org}} \citep[LCDB;][Updated 2020 October 22]{Warner2009} of which only 4930 have unambiguous rotation periods, i.e., they have a ``U'' or ``quality'' code in the LCDB of either 3 or 3-. Only 97 of those 4930 asteroids have rotation periods $>$100 hours and only 6 of those have periods $>$1000 hours. The longest\footnote{The LCDB also reports rotation periods of 1561, 1584, 1641, and 1880 hours for asteroids numbered 2440, 4892, 846, and 162058 respectively but they all have ``U'' codes of 2 so some ambiguity in those periods could still exist.} unambiguous period recorded in the LCDB currently is $1343\pm 5$  hours ($\sim$56 days) for asteroid (2056) Nancy (also known as A909 TB) \citep{Polakis2019}. Although the number of known super-slow rotators is very low, it has been suggested by \cite{Molnar2018} and \cite{Polnar2020} that the number of super-slow rotators in the LCDB is likely underestimated.  

Extracting super long periods  (P$_{rot}\gtrsim$1000 hours or $\gtrsim$40 days) from photometric observations is intrinsically difficult because long observing windows on the order of several weeks are required to fully resolve the period. However, for very long rotation periods such as these,  high-cadence observing is not critical and only helpful for eliminating potential ambiguity in the period. Low-cadence all-sky surveys that potentially observe the same object once (or more) per night over several weeks or months are therefore well suited to extracting rotation periods $\gtrsim 40$ days. Two such surveys are the Asteroid Terrestrial-impact Last Alert System (ATLAS) and the Zwicky Transient Facility (ZTF). 

The all-sky survey of ATLAS\footnote{\url{https://fallingstar.com}} \citep{Tonry2018a}
has been operating since 2015 and has made a total of $\sim$59 million individual observations of over 200\,000 known but untargeted asteroids. Many of those asteroids have been observed four times per night for several weeks during an apparition. For many asteroids there are now several apparitions worth of data. ATLAS also surveys in two custom broadband filters so a single colour for most asteroids can be extracted from the dataset, which makes crude taxonomic analysis possible \citep{Erasmus2020}.

The synoptic survey of ZTF\footnote{\url{https://www.ztf.caltech.edu}} \citep{Smith2014} achieved first light in 2017 and since full-scale operations began in 2018 has observed more than 250\,000 untargeted asteroids, over an extended period of several weeks per apparition. ZTF  currently surveys in two filters ($g$ and $r$) so a single colour is also available for most objects observed, which can assist with asteroid taxonomic classification. 

Many asteroid rotation periods have already been extracted from the ATLAS photometry data set. \cite{Durech2020} report the shape-models (and rotation periods) of $\sim$2750 asteroids of which a few had periods around 800-900 hours and the longest period found was 1236 hours. \cite{Erasmus2020} and \cite{McNeill2021} report the colours and rotation periods of >$1000$ main-belt family members and $\sim$ 40 Jupiter Trojans, respectively, of which the longest period found out of those two studies was 165 hours. Here we present photometry for a smaller number of asteroids (39), some  observed for over 5 years in the ATLAS and ZTF surveys. However, each of these 39 asteroids has a rotation period greater than 1000~hours (min: 1013~hours; median: 2390~hours; max: 4812~hours), meaning that we have discovered a class of ``super slow'' rotating asteroids that has never been seen before.

In Section~\ref{sec:observations} we describe the observations and data from these two all-sky surveys. In Section~\ref{sec:methods} we present our methodology for extracting rotation periods and other parameters for each asteroid.
Section~\ref{sec:results} presents our results and in Section~\ref{subsec:per_val} we describe our extensive period validation approaches to confirm these long period solutions.  In Section~\ref{sec:discussion} we discuss implications of our study. 

\section{Observations and data processing}
\label{sec:observations}

\subsection{All-sky surveys: ATLAS and ZTF}

Primarily designed and operated for discovering near-Earth asteroids (NEAs) on impacting orbits with Earth\footnote{As of writing ATLAS has discovered 616 NEAs of which 67 have been classified as potentially hazardous asteroids.}, ATLAS's two 0.5-m telescopes also generate accurate photometry of all detected sources in the observed fields \citep{Tonry2018b}. This photometry has already been used in several minor-body investigations like a NEA population impact-risk study \citep{Heinze2021}, the characterisation of main-belt asteroids \citep{Erasmus2020,Durech2020,Mahlke2021}, and a comparison of the shape distribution of the two Jupiter Trojan clouds \citep{McNeill2021}. The ATLAS photometry data is also bountiful and therefore serviceable for studies of transient phenomena in other astronomical disciplines \citep{Heinze2018, Smith2020}
as well as studies of the sidereal sky.

ATLAS surveys the sky in two custom filters, a ``cyan'' or \textit{c}-filter ($\sim g+r$) with a bandpass between 420--650\,nm and an ``orange'' or \textit{o}-filter ($\sim r+i$) with a bandpass between 560--820\,nm. Observations are made as 30-sec exposures for both filters where roughly 30\% of all observations are in $c$- and the remaining 70\% in the $o$-band. The approximate 5-sigma limiting magnitude per exposure is 19.7 for both filters.

ZTF is a wide-field sky survey with a $\sim$1.2-m aperture that is  designed for high-cadence time-domain astronomy \citep{Masci2019}.
ZTF provides a diverse transient astronomy community with valuable discovery alerts \citep{Patterson2019}, 
including moving targets \citep{Ye2019a,Duev2019}.
Some individual and unique asteroids have been investigated and/or discovered using ZTF photometry data \citep{Ye2019b,Ip2020,Bolin2021} and the dataset as a whole has also been exploited for a Jupiter Trojan investigation \citep{Schemel2021}. 

ZTF currently surveys the sky in two Sloan filters (\textit{g} and \textit{r}) with a third \textit{i}-filter planned for future operation.  Observations are made as 30-sec exposures with a single visit in each filter per footprint per night. The 5-sigma limiting magnitude per exposure is $\sim$20.5 \citep{Bellm2017}.

Part of the ZTF operational mode is to act as a precursor for the Legacy Survey of Space and Time (LSST), a ten-year all-sky survey that will commence in 2022/2023.
ZTF therefore provides alerts on detections of transient sources through a Kafka stream. ZTF labels known asteroids that are detected as transient sources. ANTARES is an alert broker that is designed to ingest transient alert streams \citep{Matheson2020} and is receiving and processing the ZTF detections. ANTARES is not designed to process moving objects, but does rebroadcast all moving objects in a dedicated Kafka stream that our team receives and processes. Our larger project ---- SNAPS, the Solar System Notification Alert Processing System --- will be presented in a forthcoming paper and will include some 20 000 asteroid lightcurves, the vast majority of which are unremarkable in terms of their period and amplitude.

\subsection{Processing asteroid observations}

As a first pass, we identified super-slow rotating candidates in the ATLAS and ZTF datasets independently and then later combined the data of the most promising candidates for further analysis (see Section \ref{sec:methods}). 

To distill the ATLAS and ZTF datasets we carried out the following steps on the ATLAS and ZTF 
Solar System absolute magnitudes in the four filters (i.e., H$_{c}$, H$_{o}$, H$_{g}$, and H$_{r}$) that are generated from the observed magnitudes using the  H-G model with a generic G-parameter of 0.15:
\begin{enumerate}
 \item For both the H$_{c}$ and H$_{o}$ ATLAS data for each object we discarded outlying data points by only retaining data where the H value lay within 1.0 magnitude of the median H value for that specific object. We also removed poorer quality data by rejecting data points where the uncertainty in the observed magnitude was $>$0.1 magnitude. For ZTF, we discarded observations with a Real-Bogus score $<0.6$ (\cite{Duev2019b}).
\item To limit our study to objects with a high confidence in extracted rotation period we also only considered objects with a light-curve amplitude $\gtrsim$0.4 mag and therefore also only pre-selected objects where the standard deviation of the ``scrubbed'' absolute magnitude data was $>$0.2 magnitude.
 \item We also only retained data for asteroids where we had at least 150 and 50 number of data points in the \textit{o-} and \textit{c-}filter respectively for ATLAS and at least 50 combined data points in the \textit{g} and \textit{r} filters for ZTF.
\end{enumerate}

The result of these selection criteria distilled the $\sim$250\,000 objects 
in each of ATLAS and ZTF databases down to around $\sim$10\,000 asteroids for which we extracted lightcurve solutions from the  ATLAS and ZTF data independently with the \texttt{SciPy} \citep{SciPy2020} implementation of Lomb-Scargle \citep[LS;][]{Lomb1976,Scargle1982}, searching for periods between 1-10\,000 hours.
From this list, we identified a few hundred objects in total where the LS analysis best fits showed potential very long periods ($>$1000~hours) in either the ATLAS or ZTF data (or both).
Examples of our data processing are shown in Figures~\ref{fig:example_app_phase} and~\ref{fig:example_abs_per_fold}.

After visible inspection of the periodograms and folded light-curves we identified a first release of
objects for which we had both ATLAS and ZTF data and where both data sets independently showed a convincing indication of very slow rotation ($>$1000~hours). Using the final candidate list we performed more refined period and colour analysis on the combined ATLAS and ZTF data for these objects (see Section \ref{sec:methods}).  

\section{Methods}
\label{sec:methods}

\subsection{Phase Curve Parameters, Colours, and Refined Rotation Periods}
\label{subsec:rot_per_col_phase} 

\begin{figure*}
\centering
\subfloat[]{
  \includegraphics[width=0.49\textwidth]{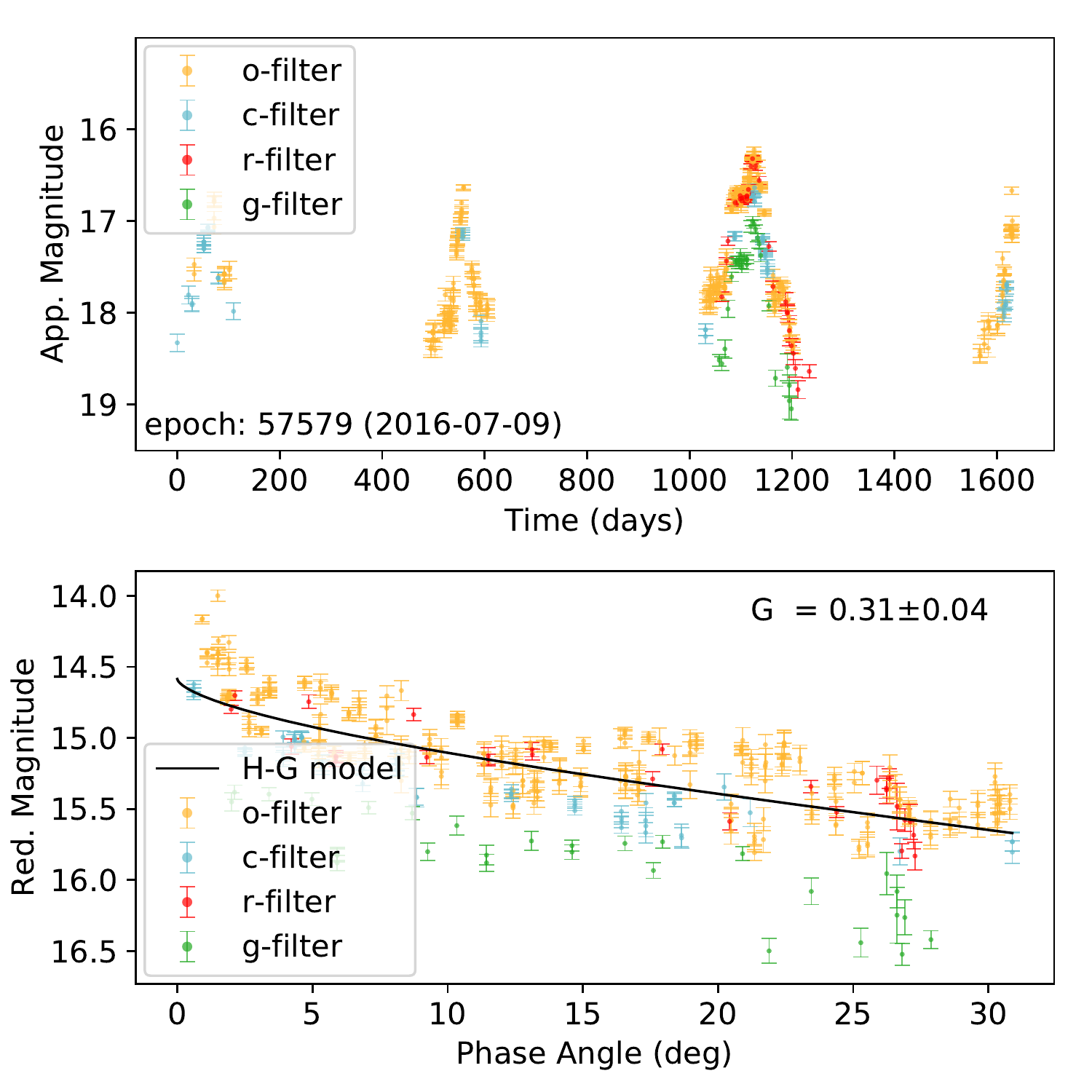}}
\subfloat[]{
  \includegraphics[width=0.49\textwidth]{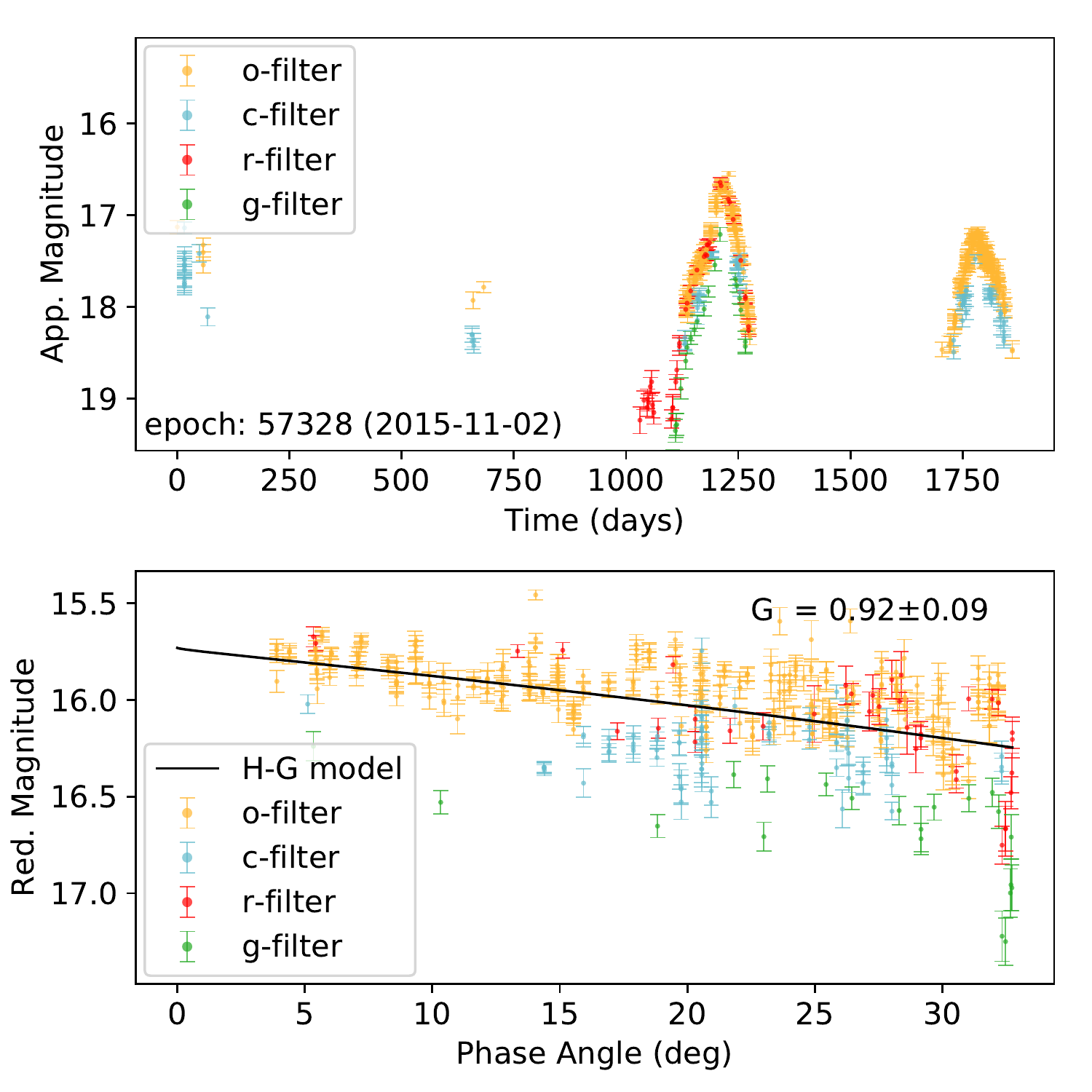}}
  
\caption{Example ATLAS ($c$- and $o$-filter) and ZTF ($g$- and $r$-filter) data for  (a) asteroid (16409) 1986 CZ1 and (b) asteroid (52534) 1996 TB15. (Top) The raw survey data in apparent observed magnitude showing data from four apparitions for the specific objects. (Bottom) Reduced magnitude with the observer-to-object distance-dependence on the brightness removed. The H-G fitted model is shown, with the derived G value shown in the top-right corner. \textit{Equivalent plots and data for all 39 objects listed in Table \ref{tab:data} will be available online.}}
\label{fig:example_app_phase}  
\end{figure*}

Before extracting refined rotation periods we first recalculate more accurate absolute magnitudes by refitting the phase curve on the combined ATLAS and ZTF apparent magnitude data with the H-G model formulated by \citet{Bowell1989} and summarised in \citet{Dymock2007}. An example of this is shown in the bottom panel of Figure \ref{fig:example_app_phase}. 

\begin{figure*}
\centering
\subfloat[]{
  \includegraphics[width=0.49\textwidth]{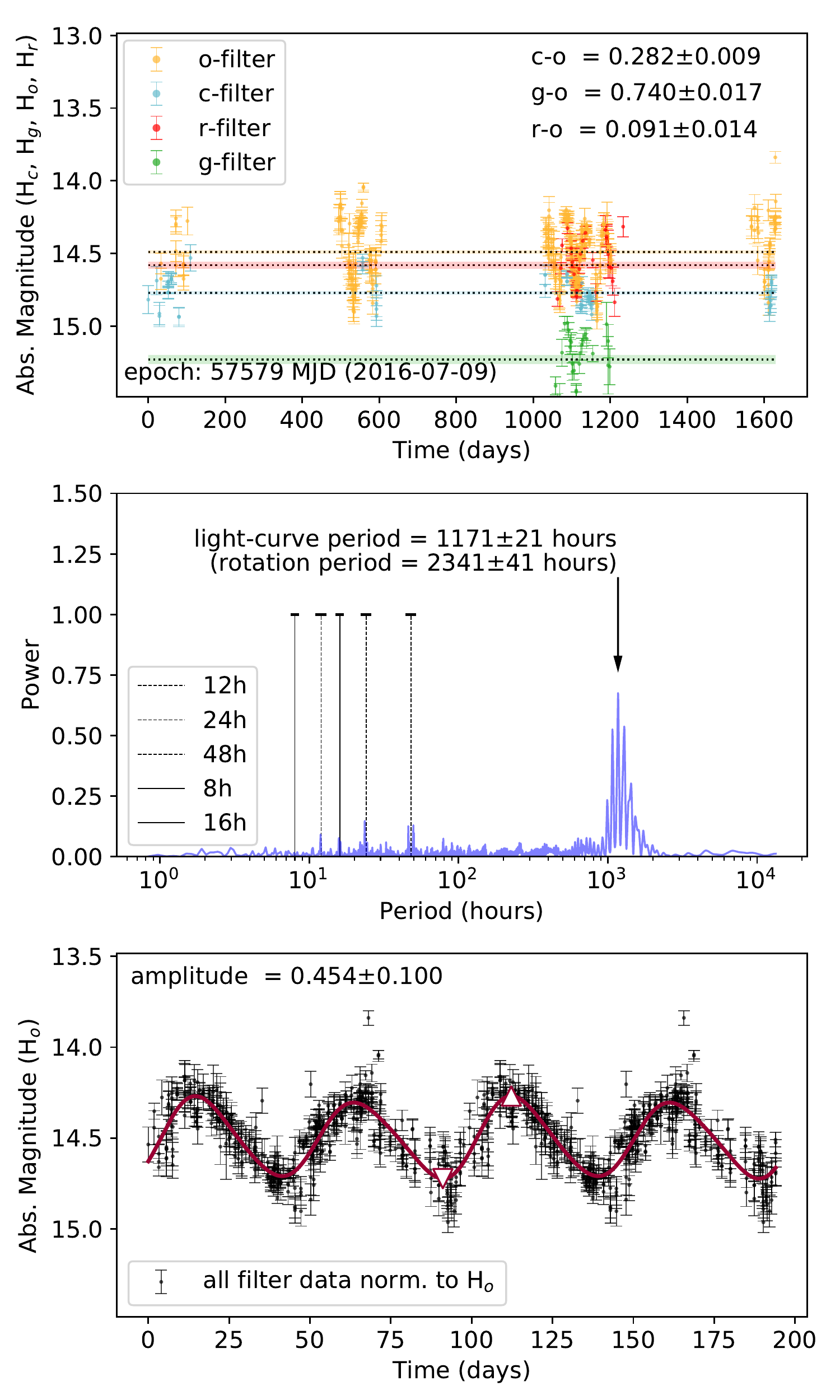}}
\subfloat[]{
  \includegraphics[width=0.49\textwidth]{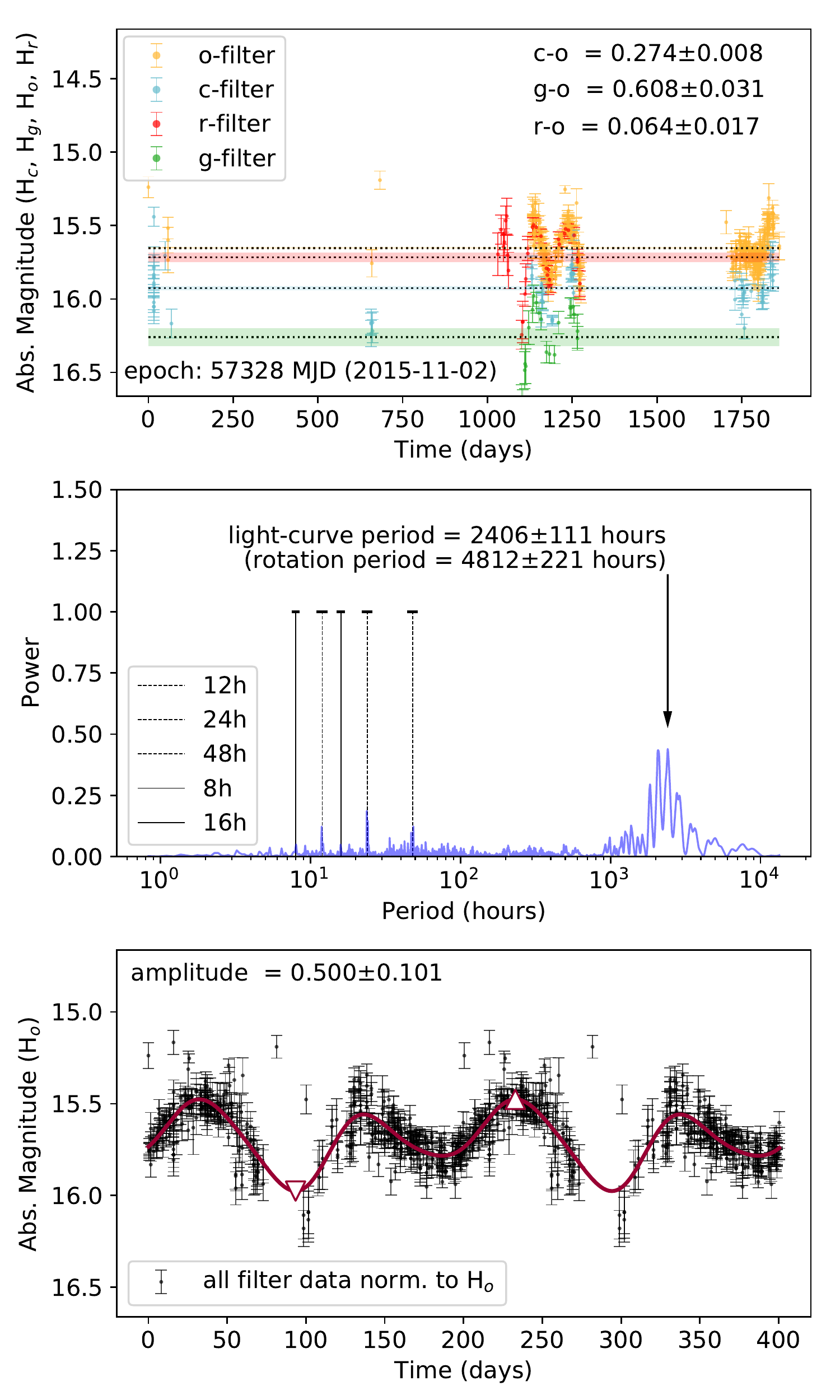}}
  
\caption{Example ATLAS ($c$- and $o$-filter) and ZTF ($g$- and $r$-filter) data for (a) asteroid (16409) 1986 CZ1 and (b) asteroid (52534) 1996 TB15. (Top) Absolute magnitudes H$_{c}$, H$_{o}$, H$_{g}$ and H$_{r}$ as a function of time. The colours are calculated using the median magnitude in each filter (see Section \ref{sec:methods}) and shown in the top-right corner. (Middle) The Lomb-Scargle periodogram of the combined filter data (see Section \ref{sec:methods}). The arrow shows the strongest periodogram peak with the light-curve period (i.e., half the rotation period) labeled, in the case of asteroid (52534) 1996 TB15 it is 4812 hours ($\sim$200 days) which is the longest period reported in this study.  Peaks at 8- 12- 16- 24- and 48-hours (positions indicated with vertical lines) are almost always present in the ATLAS/ZTF data periodograms because of the strong 1-day cadence of both surveys (see Section \ref{subsec:per_val} for further detail).
(Bottom) The combined filter data (from the derived colours shown in the top panel) folded with the best solution (shown in the middle panel)
and with a Fourier series fit shown as a solid line. The white triangles show the minimum and maximum values of the fit and are used to determine the light-curve amplitude displayed in the top-left corner of the panel. \textit{Equivalent plots for all 39 objects listed in Table \ref{tab:data} will be available online.}}
\label{fig:example_abs_per_fold}
\end{figure*}

With the new refined absolute magnitudes we then calculate \textit{c-o} and \textit{g-r} colours. The colours are calculated by determining the median H$_{c}$, H$_{o}$, H$_{g}$ and H$_{r}$ magnitudes of a randomly selected 50\% subset of the \textit{c}-, \textit{o}-, \textit{g}-, and \textit{r}-filter data, respectively, and repeating that process 10 times and calculating the average. The final colour is defined as the difference between the average of the median magnitudes. The uncertainty in the colour value incorporates the standard deviation of the median magnitudes in the previous step. An example of this is shown in the top panel of Figure \ref{fig:example_abs_per_fold} with the calculated colour the distance between the respective dotted horizontal lines and the standard deviation of the 10 median magnitudes shown as shaded regions. The colour and uncertainties in colour for each object are tabulated in Table \ref{tab:data}. 

To determine refined rotation periods we combine all filter data  by using the $o$-filter data as reference (because it was the most abundant data) and subtracting the relevant colours from the relevant filter data. We repeat our search for periods between 1-10\,000 hours and compute a periodogram of each object on the combined data (see example periodogram in the middle panel of Figure \ref{fig:example_abs_per_fold}). From these more refined LS periodograms we made a final selection of 39 objects for which we have a high confidence that the rotation period is more than 1000 hours. The extracted rotation periods (twice the light-curve period) and the uncertainty in the rotation period of those 39 super-slow rotating asteroids are recorded in Table \ref{tab:data}.

\subsection{Taxonomic Determination}
\label{subsec:tax}

\begin{figure}
	\includegraphics[width=\columnwidth]{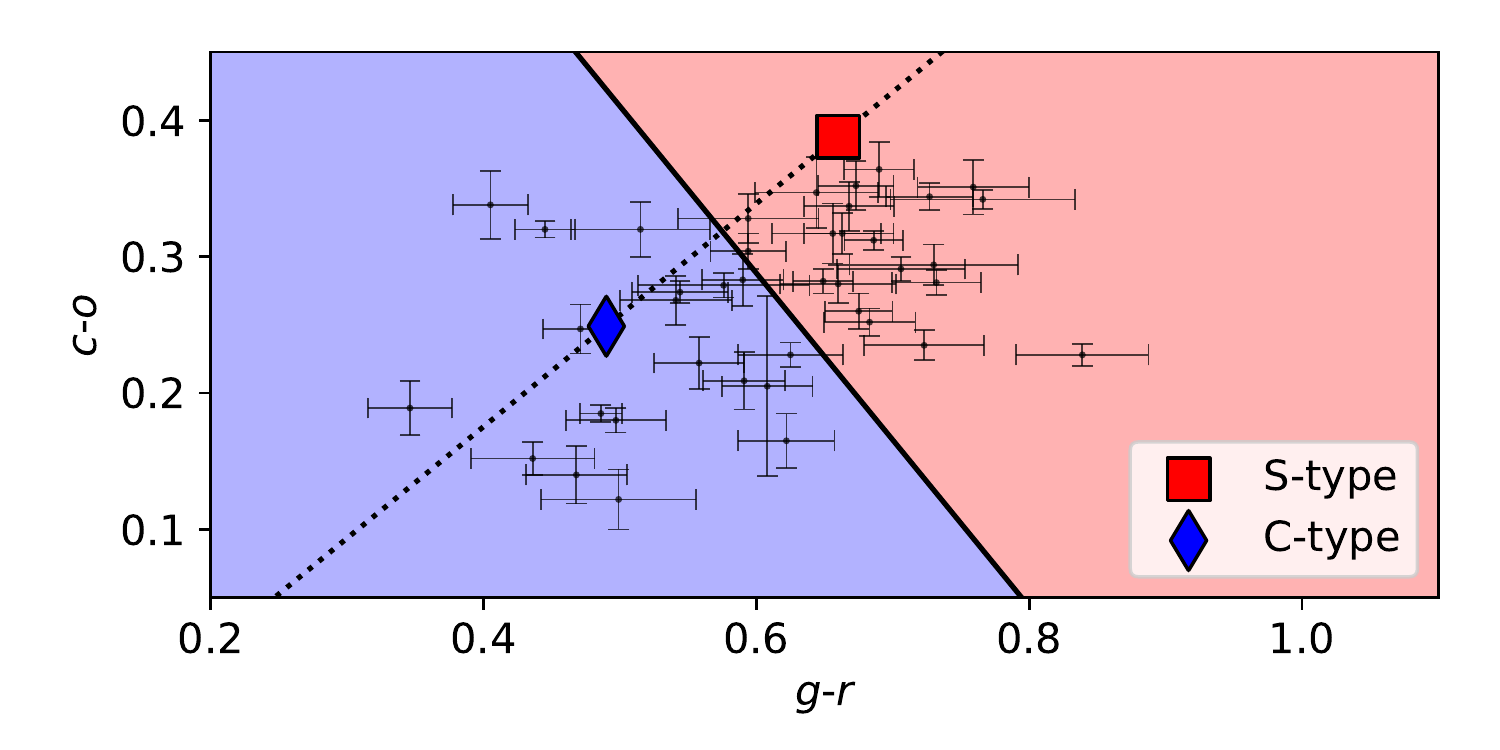}
    \caption{Colour-colour plot showing the \textit{g-r} and \textit{c-o} colours on the x- and y-axis respectively. Indicated with a red square and a blue diamond symbol is the expected colour values for a typical S- and C-type asteroid using \textit{c-o} values reported in \citet{Erasmus2020} and estimated \textit{g-r} values from Figure 8 in \citet{Ivezic2001}. The solid line shows a decision boundary between S- and C-like asteroids and used for calculating the taxonomic probability of each object (see Section \ref{subsec:tax} and probabilities in Table \ref{tab:data}).}
    \label{fig:color_color}
\end{figure}

To determine a crude taxonomy for each of the 39 objects we report here, we make use of the $c-o$ and $g-r$ colours. Plotted in Figure \ref{fig:color_color} are the $c-o$ and $g-r$ colours from Table \ref{tab:data} with the expected $g-r$ and $c-o$ colour for a typical S-type and C-type asteroid indicated with a red square symbol and a blue diamond symbol respectively \citep[values obtained from ][]{Ivezic2001,Erasmus2020}. We determine the taxonomic probability of each object using the measured colour and colour uncertainty with respect to the S/C-type decision boundary shown as a solid black line in Figure \ref{fig:color_color}. This is a similar approach to that used in \citet[][see Figure 4-d]{Erasmus2017}. The taxonomic probabilities are tabulated in Table \ref{tab:data}.

\section{Results}
\label{sec:results}

The results of our investigation (periods, amplitudes, colours, and taxonomic probabilities) for the 39 objects we identify as super slow rotators from the ATLAS and ZTF datasets are summarised in Table \ref{tab:data} below. 
These asteroids all have rotation periods greater than 1000~hours (42~days).
32~of these objects have periods longer than any previously reported unambiguous rotation periods currently in the Asteroid Light Curve Database. 
Our data selection criteria (see Section \ref{sec:observations}) distilled the ATLAS and ZTF data sets down to roughly 10 000 objects from which we identify 39 super-slow rotators. From our final list we therefore conclude that super-slow rotators must be at least 0.4\% of the main belt asteroid population.

The longest rotation period that we report here is 4812~hours (200~days). The observing season for any given asteroid is rarely much longer than this, so we are biased against detecting rotation periods longer than 6~months or so. In other words, the distribution of asteroid rotation periods may extend (far) beyond the longest period reported here. However, detecting such extremely long periods will be difficult with all current and future surveys.


Our taxonomic determination of the 39 super-slow rotators (see Figure \ref{fig:color_color} and Table \ref{tab:data}) reveals that there is roughly equal probability (51\% vs. 48\%) of discovering a S- or C-like slow rotator with a fairly uniform spread in colours, which suggests that slow rotation is most likely not favoured by a specific taxonomic class.

We  also include other values in the table such as the albedo and semi-major axis of each object that we obtained from JPL Small-Body Database.  
The ranges of semi-major axis for the objects reported here also span roughly uniformly between 2.1 and 3.25 AU (i.e., the entire main-belt) so it also does not appear that slow-rotators preferentially found in any location in the main belt. The albedos show that we also have very dark (as low as 3\% reflective) and very bright (as high as 46\% reflective) asteroids on our list so there is also no correlation between albedo and a super-slow rotating status.

\begin{landscape}
    \begin{table}
	\centering
	\caption{Super-slow rotating asteroids}
	\begin{tabular}{|l|l|cc|cc|c|cc|cc|cc|c|c|c|} 
		\hline
		No. & Object & H$^{1}_{V}$ & H$^{2}_{c}$ &\multicolumn{2}{c|}{\# Data Points} & Obs. &\multicolumn{2}{c|}{Colours} & Rotation & Light-Curve & C-like & S-like & Albedo$^1$ & $a^1$ & G\\
		& & & & ATLAS & ZTF & Window & $c-o$ & $g-r$ & Period & Amplitude &\multicolumn{2}{c|}{Probability}& & &\\
		& & (mag) & (mag)&  &  & (days) &(mag) & (mag) & (hours) & (mag) &(\%) & (\%)& & (au) & \\
		\hline
	    01&1183 Jutta (1930 DC)&12.4&12.1&803&35&1977&0.18$\pm$0.01&0.50$\pm$0.04&3712$\pm$0101&0.46$\pm$0.19&100&0&0.045&2.384&0.07$\pm$0.02\\
02&2339 Anacreon (2509 P-L)&13.2&13.1&748&31&1697&0.29$\pm$0.01&0.71$\pm$0.05&3608$\pm$0065&0.63$\pm$0.15&0&100&0.121&2.526&0.15$\pm$0.03\\
03&2831 Stevin (1930 SZ)&12.7&12.6&735&26&1726&0.29$\pm$0.01&0.73$\pm$0.06&3213$\pm$0207&0.50$\pm$0.22&0&100&0.651&2.226&0.20$\pm$0.03\\
04&2862 Vavilov (1977 JP)&12.8&12.9&1138&46&1885&0.34$\pm$0.01&0.77$\pm$0.07&3015$\pm$0051&0.81$\pm$0.14&0&100&0.260&2.201&0.36$\pm$0.04\\
05&3412 Kafka (1983 AU2)&13.4&13.5&798&59&1753&0.28$\pm$0.01&0.73$\pm$0.03&2766$\pm$0040&0.63$\pm$0.13&0&100&0.231&2.225&0.29$\pm$0.03\\
06&3884 Alferov (1977 EM1)&13.2&12.9&258&59&1713&0.19$\pm$0.02&0.35$\pm$0.03&2495$\pm$0073&0.68$\pm$0.14&100&0&0.059&3.112&0.01$\pm$0.03\\
07&5179 Takeshima (1989 EO1)&13.7&14.0&463&110&1979&0.26$\pm$0.01&0.67$\pm$0.02&1069$\pm$0009&0.70$\pm$0.16&0&100&0.416&2.311&0.51$\pm$0.08\\
08&6726 Suthers (1991 PS)&14.0&14.2&467&91&1958&0.31$\pm$0.01&0.69$\pm$0.02&4204$\pm$0150&0.53$\pm$0.10&0&100&0.207&2.287&0.29$\pm$0.04\\
09&9892 (1995 YN3)&14.1&14.4&557&84&1689&0.34$\pm$0.01&0.73$\pm$0.03&2315$\pm$0033&0.70$\pm$0.13&0&100&0.460&2.264&0.39$\pm$0.06\\
10&10314 (1990 RF)&12.1&12.0&920&76&1820&0.18$\pm$0.01&0.49$\pm$0.02&1557$\pm$0014&0.35$\pm$0.11&100&0&0.047&3.218&0.12$\pm$0.02\\
11&13423 Bobwoolley (1999 VR22)&13.2&13.1&531&44&1529&0.15$\pm$0.01&0.44$\pm$0.05&3760$\pm$0088&0.77$\pm$0.15&100&0&0.076&2.749&0.19$\pm$0.04\\
12&13999 (1993 FH43)&14.7&14.9&138&56&1672&0.21$\pm$0.02&0.59$\pm$0.03&1330$\pm$0009&0.43$\pm$0.13&100&0&0.224&2.333&0.39$\pm$0.07\\
13&14979 (1997 TK1)&14.2&14.3&229&68&1893&0.35$\pm$0.02&0.67$\pm$0.03&4440$\pm$0105&0.61$\pm$0.12&0&100&0.225&2.648&0.26$\pm$0.06\\
14&16409 (1986 CZ1)&14.5&14.8&374&57&1630&0.28$\pm$0.01&0.65$\pm$0.02&2341$\pm$0041&0.45$\pm$0.10&0&100&-&2.179&0.31$\pm$0.04\\
15&17582 (1994 WL)&13.8&14.0&386&56&1998&0.25$\pm$0.01&0.68$\pm$0.03&1209$\pm$0018&0.45$\pm$0.13&0&100&0.395&2.360&0.40$\pm$0.05\\
16&18096 (2000 LM16)&13.9&13.7&299&71&1652&0.23$\pm$0.01&0.62$\pm$0.04&3626$\pm$0142&0.64$\pm$0.12&81&19&0.119&2.404&0.06$\pm$0.04\\
17&19818 Shotwell (2000 SB150)&13.6&13.5&663&67&1687&0.23$\pm$0.01&0.72$\pm$0.04&2368$\pm$0037&0.83$\pm$0.17&0&100&0.261&2.365&0.26$\pm$0.04\\
18&24412 Ericpalmer (2000 AM243&12.8&12.8&567&54&1870&0.32$\pm$0.01&0.45$\pm$0.02&1269$\pm$0008&0.35$\pm$0.13&100&0&0.040&3.149&0.06$\pm$0.03\\
19&24825 (1995 QB2)&14.5&14.4&294&71&1754&0.28$\pm$0.01&0.66$\pm$0.04&1248$\pm$0013&0.54$\pm$0.19&0&100&0.176&2.237&0.18$\pm$0.04\\
20&25629 Mukherjee (2000 AH52)&15.3&15.5&124&74&1178&0.22$\pm$0.02&0.56$\pm$0.03&4141$\pm$0649&0.76$\pm$0.12&100&0&0.244&2.238&0.31$\pm$0.08\\
21&32853 Dobereiner (1992 SF2)&13.7&14.2&545&44&1582&0.28$\pm$0.01&0.58$\pm$0.06&2571$\pm$0047&0.69$\pm$0.15&75&25&0.247&2.537&0.61$\pm$0.08\\
22&36707 (2000 RP29)&14.0&13.7&167&79&1515&0.32$\pm$0.02&0.66$\pm$0.04&1013$\pm$0035&0.76$\pm$0.18&0&100&-&2.782&0.16$\pm$0.09\\
23&40470 (1999 RN47)&14.6&14.5&179&74&1601&0.14$\pm$0.02&0.47$\pm$0.04&1399$\pm$0068&0.62$\pm$0.16&100&0&0.172&2.475&0.12$\pm$0.04\\
24&45212 (1999 XP180)&14.3&14.4&276&105&1639&0.32$\pm$0.01&0.66$\pm$0.03&1741$\pm$0159&0.71$\pm$0.13&0&100&-&2.570&0.38$\pm$0.10\\
25&46332 (2001 QD276)&13.5&13.5&248&75&1068&0.34$\pm$0.02&0.67$\pm$0.03&2010$\pm$0033&0.65$\pm$0.15&0&100&0.118&3.057&0.10$\pm$0.07\\
26&48365 (3106 T-2)&14.8&14.5&134&53&1577&0.33$\pm$0.02&0.59$\pm$0.05&1611$\pm$0014&0.76$\pm$0.13&24&76&0.148&2.592&0.20$\pm$0.08\\
27&52534 (1996 TB15)&15.2&16.0&404&57&1861&0.27$\pm$0.01&0.54$\pm$0.04&4812$\pm$0221&0.50$\pm$0.10&100&0&-&1.923&0.92$\pm$0.09\\
28&57107 (2001 OF74)&15.8&15.5&95&81&1677&0.36$\pm$0.02&0.69$\pm$0.03&1124$\pm$0065&0.44$\pm$0.14&0&100&-&2.321&-0.03$\pm$0.05\\
29&58659 (1997 WZ57)&15.5&15.6&139&65&1672&0.35$\pm$0.02&0.76$\pm$0.04&4624$\pm$1500&0.94$\pm$0.07&0&100&0.203&2.383&0.32$\pm$0.09\\
30&63354 (2001 FU140)&14.1&14.1&103&54&617&0.27$\pm$0.02&0.54$\pm$0.04&1592$\pm$0118&0.70$\pm$0.13&100&0&-&3.119&0.34$\pm$0.11\\
31&71727 (2000 HB25)&13.3&13.1&259&107&1779&0.25$\pm$0.02&0.47$\pm$0.03&2158$\pm$0053&0.54$\pm$0.14&100&0&0.049&3.116&0.05$\pm$0.04\\
32&86019 (1999 JG127)&14.5&14.0&263&64&1129&0.17$\pm$0.02&0.62$\pm$0.04&2432$\pm$0048&0.68$\pm$0.18&100&0&0.230&2.416&0.05$\pm$0.04\\
33&87133 (2000 NY4)&15.1&15.0&119&99&1677&0.30$\pm$0.01&0.59$\pm$0.03&2414$\pm$0336&0.64$\pm$0.12&37&63&0.368&2.399&0.12$\pm$0.07\\
34&116629 (2004 BD122)&14.8&15.2&72&62&229&0.32$\pm$0.02&0.51$\pm$0.05&4638$\pm$1395&1.14$\pm$0.13&99&1&-&3.116&0.94$\pm$0.26\\
35&119428 (2001 TP120)&14.9&14.9&279&97&1133&0.28$\pm$0.02&0.59$\pm$0.03&4121$\pm$0136&0.52$\pm$0.16&73&27&0.303&2.353&0.30$\pm$0.08\\
36&143088 (2002 XT9)&15.3&15.1&71&82&1515&0.20$\pm$0.07&0.61$\pm$0.03&1461$\pm$0084&0.72$\pm$0.14&95&5&0.313&2.650&0.20$\pm$0.08\\
37&172608 (2003 WA86)&15.0&15.2&121&77&199&0.35$\pm$0.03&0.64$\pm$0.05&1554$\pm$0172&0.81$\pm$0.11&0&100&0.185&2.559&0.31$\pm$0.12\\
38&310440 (2000 BL2)&16.4&16.1&88&84&220&0.34$\pm$0.03&0.41$\pm$0.03&1980$\pm$0525&0.47$\pm$0.11&100&0&0.043&2.539&0.04$\pm$0.07\\
39&311760 (2006 TO96)&15.9&15.4&22&55&141&0.12$\pm$0.02&0.50$\pm$0.06&2600$\pm$0534&1.22$\pm$0.15&100&0&0.028&2.693&0.03$\pm$0.16\\

	    \hline
	    \multicolumn{16}{l}{$^1$ Absolute magnitude (H$_V$), albedo and semi-major axis ($a$) obtained from the JPL Small-Body Database (SBDB) via the \texttt{Astropy} affiliated package \texttt{sbpy} \citep{Mommert_2019}.  }\\
	    \multicolumn{16}{l}{$^2$ H$_c$ is derived by adding the $c-o$ colour to H$_o$ which is obtained from the phase curve fits to the $o$-filter data shown in Figure(s)
	    \ref{fig:example_app_phase}.}\\	    
	    \multicolumn{16}{l}{$^*$ The periodogram of Jutta (see online Figure \ref{fig:example_abs_per_fold}) also has a equally strong peak at a rotation period 4076~hours. Folding the period at either 3712 or 4076~hours displays some ``dephasing'' in the folded light-curve.}\\
	    \multicolumn{16}{l}{\cite{Stephens2011a} reports a period of 212.5~hours for Jutta but we see no evidence of that in our periodogram.}\\
	    
	\end{tabular}
	\label{tab:data}
    \end{table}
\end{landscape}

\section{Period Validation}
\label{subsec:per_val}

\subsection{Follow up observations}

To validate our extracted rotation periods we performed follow-up light-curve observations of two objects, 1621 Druzhba (1930 SZ) and 2831 Stevin (1926 TM).

Druzhba was chosen as a follow-up candidate from our list of a few hundred potential super-slow rotators because the ATLAS/ZTF data produced a strong periodogram peak at a period of 2781 hours but strictly speaking the strongest periodogram peak, by a small margin, was found at  approximately 16-hours. This latter peak (as well as peaks at 8- 12- 24- and 48-hours) are almost always present in the ATLAS/ZTF data periodograms because of the strong 1-day cadence of both surveys (see vertical lines in the middle panel of Figures \ref{fig:example_abs_per_fold}, \ref{fig:Druzhba_ATLAS_ZTF}, and \ref{fig:Stevin_ATLAS_ZTF}). It is also difficult to rule out periods at those peaks using ATLAS/ZTF data because of the sparseness of the data inter-night. Folding the ATLAS-ZTF data for Druzhba at the (incorrect) long period also produced a convincing lightcurve (see bottom panel of Figure \ref{fig:Druzhba_ATLAS_ZTF}). 

Stevin was chosen as a follow-up candidate because the ATLAS/ZTF data produced the strongest periodogram peak at a period of 3213 hours but it was one of the objects with the least convincing periodogram, i.e., no distinct isolated peak at the long period. The folded data at the long period also had relatively large scatter (see Figure \ref{fig:Stevin_ATLAS_ZTF} for periodogram and folded light-curve of Stevin). 
For Stevin, our ATLAS+ZTF solution also suggested a lightcurve amplitude of at least 0.5~magnitude (and potentially as much as 1~magnitude when accounting for the large scatter). 

The purpose of the follow-up measurements was therefore
(a) to test for periods near 24~hours that would be difficult to detect in the ATLAS/ZTF datasets due to aliasing from the sparse observational cadence and (b) to confirm our long period solutions.

To make these additional measurements we used a 1-meter PlaneWave telescope at Lowell Observatory, equipped with a Finger Lakes Instruments ML16803 CCD camera.  Data were taken over $\sim$90 nights during the months of 2020-12 through 2021-03 in an `open' filter configuration (filters were not yet available for this new facility).
Each observing night typically had 25-50 observations per object per night with a cadence of approximately 10 minutes.
This high-cadence approach is complementary to the low-cadence ATLAS and ZTF datasets.

\begin{figure}
	\includegraphics[width=\columnwidth]{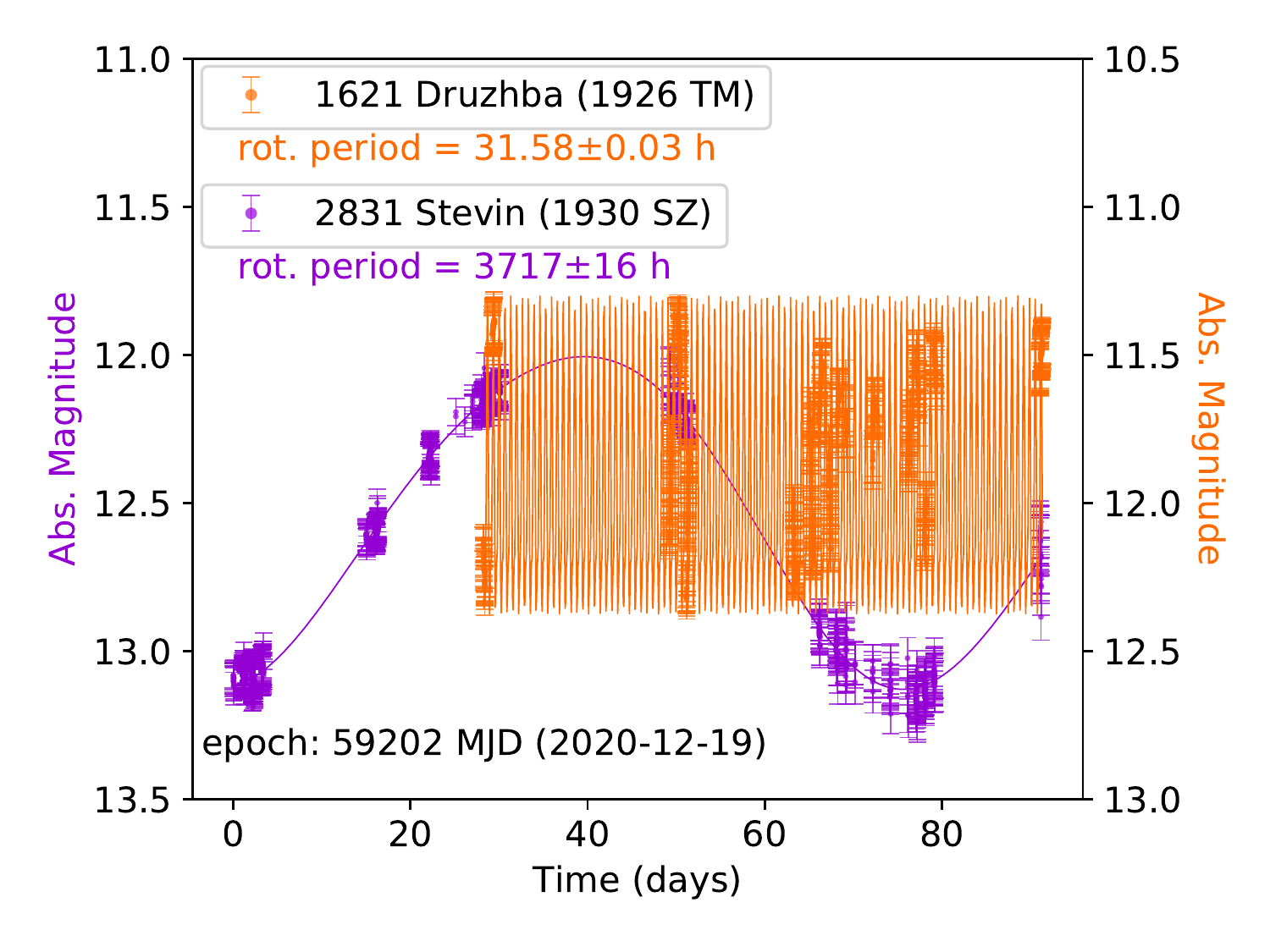}
    \caption{Follow-up observation photometry obtained (see Section \ref{subsec:per_val}) for candidate super slow rotating asteroids 1621 Druzhba (in orange)  and 2831 Stevin (in purple). These follow-up observations confirm Stevin's super-slow rotating status while Druzhba was shown to not have a super-slow rotation period.}
    \label{fig:Stevin_Druzhba_plot}
\end{figure}

Image files from the ML16803 were pre-processed with dark/flat/bias corrections using the {\tt ccdproc} software package \citep{Craig2017}.  Resulting debiased/flattened image frames were then processed by the {\tt PHOTOMETRYPIPLINE} package \citep{Mommert2017} to produce time-series photometry on the moving objects. Our follow-up observations are shown in Figure \ref{fig:Stevin_Druzhba_plot}.

\begin{figure}
	\includegraphics[width=\columnwidth]{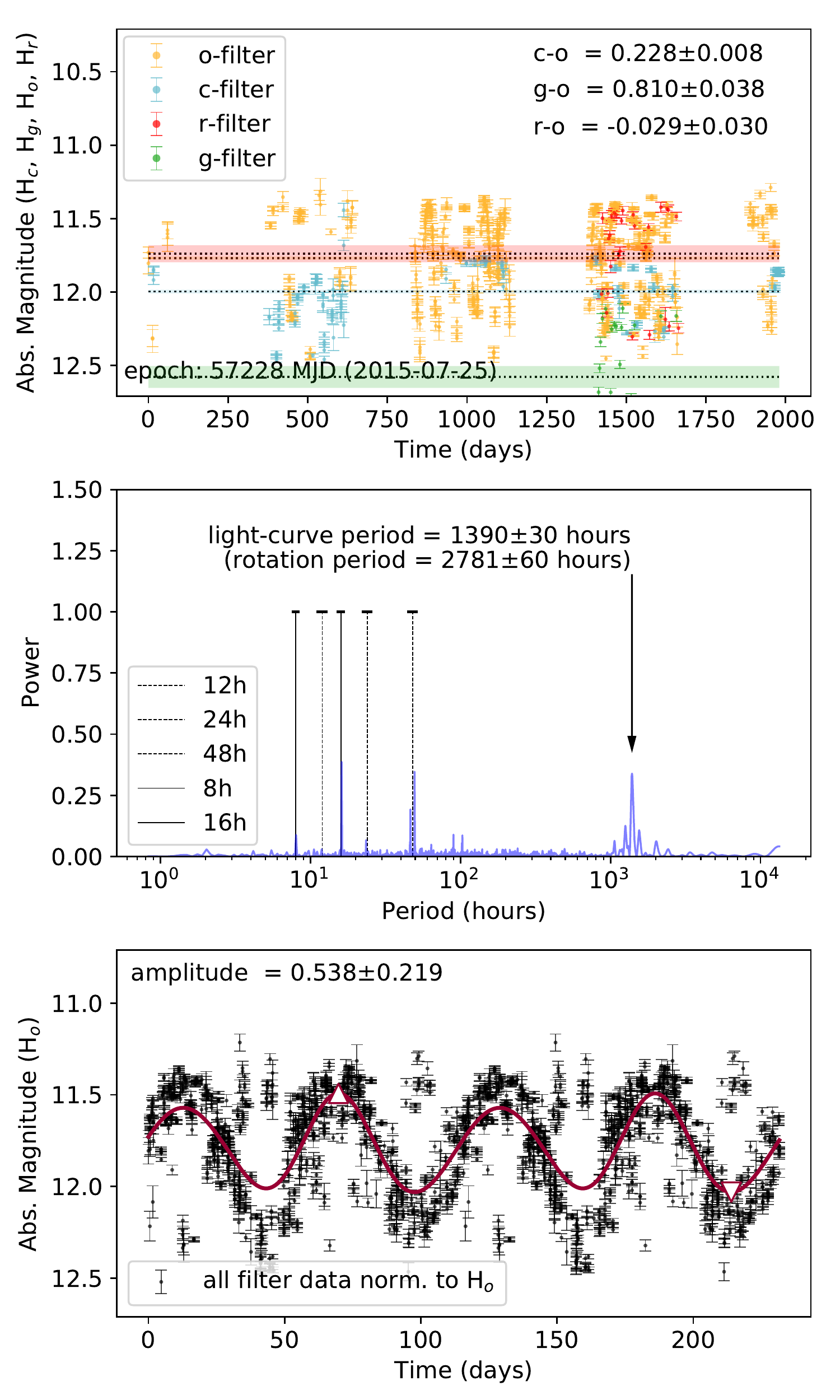}
    \caption{ATLAS ($c$- and $o$-filter) and ZTF ($g$- and $r$-filter) data for asteroid (1621) Druzhba. (Top) Absolute magnitudes H$_{c}$, H$_{o}$, H$_{g}$ and H$_{r}$ as a function of time. (Middle) The Lomb-Scargle periodogram of the combined filter data (see Section \ref{sec:methods}). The arrow shows the periodogram peak corresponding to the (incorrect, see text in Sections \ref{subsec:per_val}) rotation period of 2781~hours. The follow-up observations for Druzhba (see Figure \ref{fig:Stevin_Druzhba_plot}) show that the strongest periodogram peak at $\sim$16~hours is actually the peak that corresponds to the correct rotation period of $\sim$32~hours. 
    (Bottom) The combined filter data (from the derived colours shown in the top panel) folded with the
    incorrect 2781~hour period illustrating that a relatively clean fold can also be obtained after folding with incorrect periods (see text in Sections \ref{subsec:per_val} for more discussion).}
    \label{fig:Druzhba_ATLAS_ZTF}
\end{figure}

\begin{figure}
	\includegraphics[width=\columnwidth]{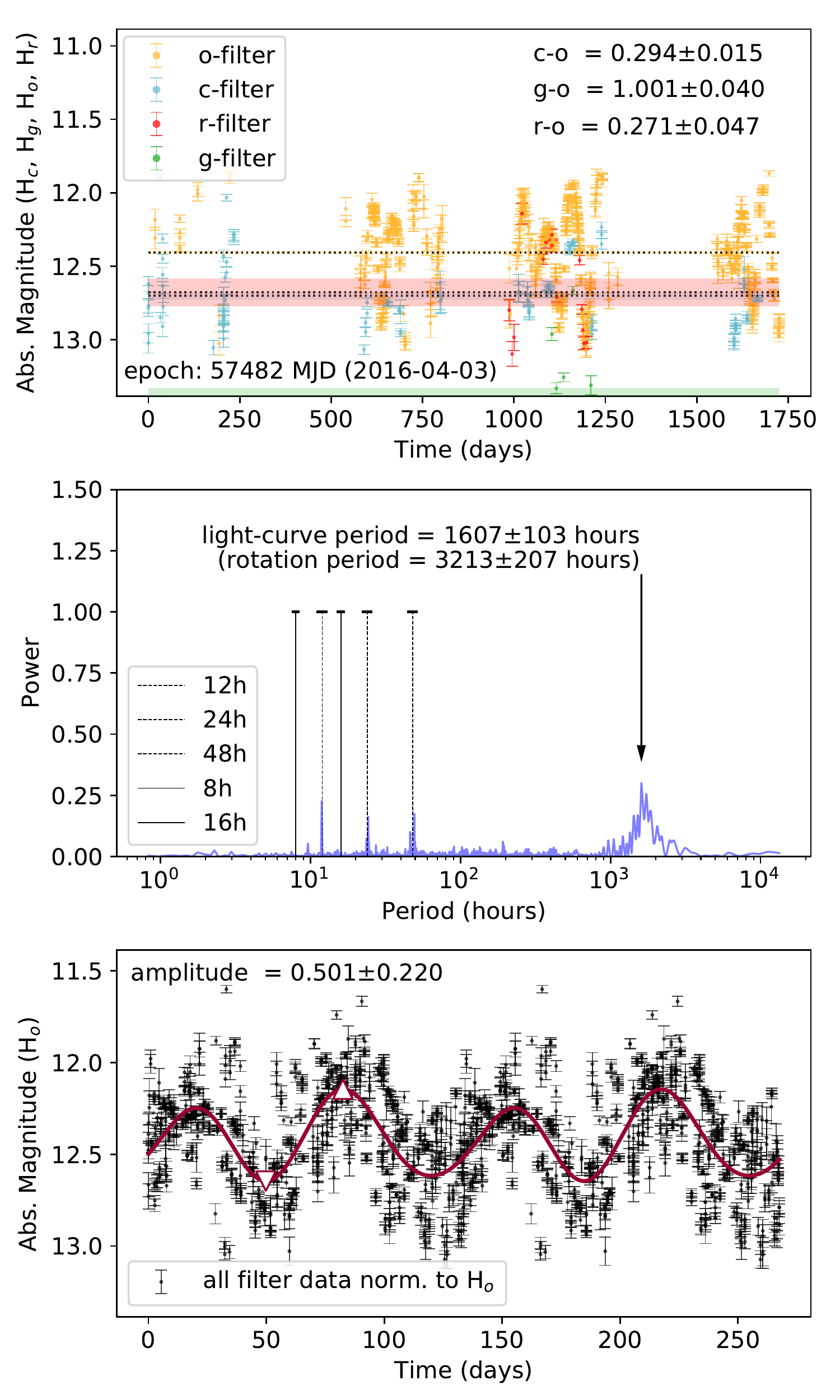}
    \caption{ATLAS ($c$- and $o$-filter) and ZTF ($g$- and $r$-filter) data for asteroid (2831) Stevin. (Top) Absolute magnitudes H$_{c}$, H$_{o}$, H$_{g}$ and H$_{r}$ as a function of time. (Middle) The Lomb-Scargle periodogram of the combined filter data (see Section \ref{sec:methods}). The arrow shows the strongest periodogram peak with the rotation period of 3213~hours labeled. The follow-up observations for Stevin (see Figure \ref{fig:Stevin_Druzhba_plot}) show a period of 3717~hours which corresponds to one of the adjacent peaks in the strongest periodogram peak (see Figure \ref{fig:Stevin_ATLAS_ZTF_alt} for data folded at this period). 
    (Bottom) The combined filter data (from the derived colours shown in the top panel) folded with the
    best solution (shown in the middle panel).}
    \label{fig:Stevin_ATLAS_ZTF}
\end{figure}

\begin{figure}
	\includegraphics[width=\columnwidth]{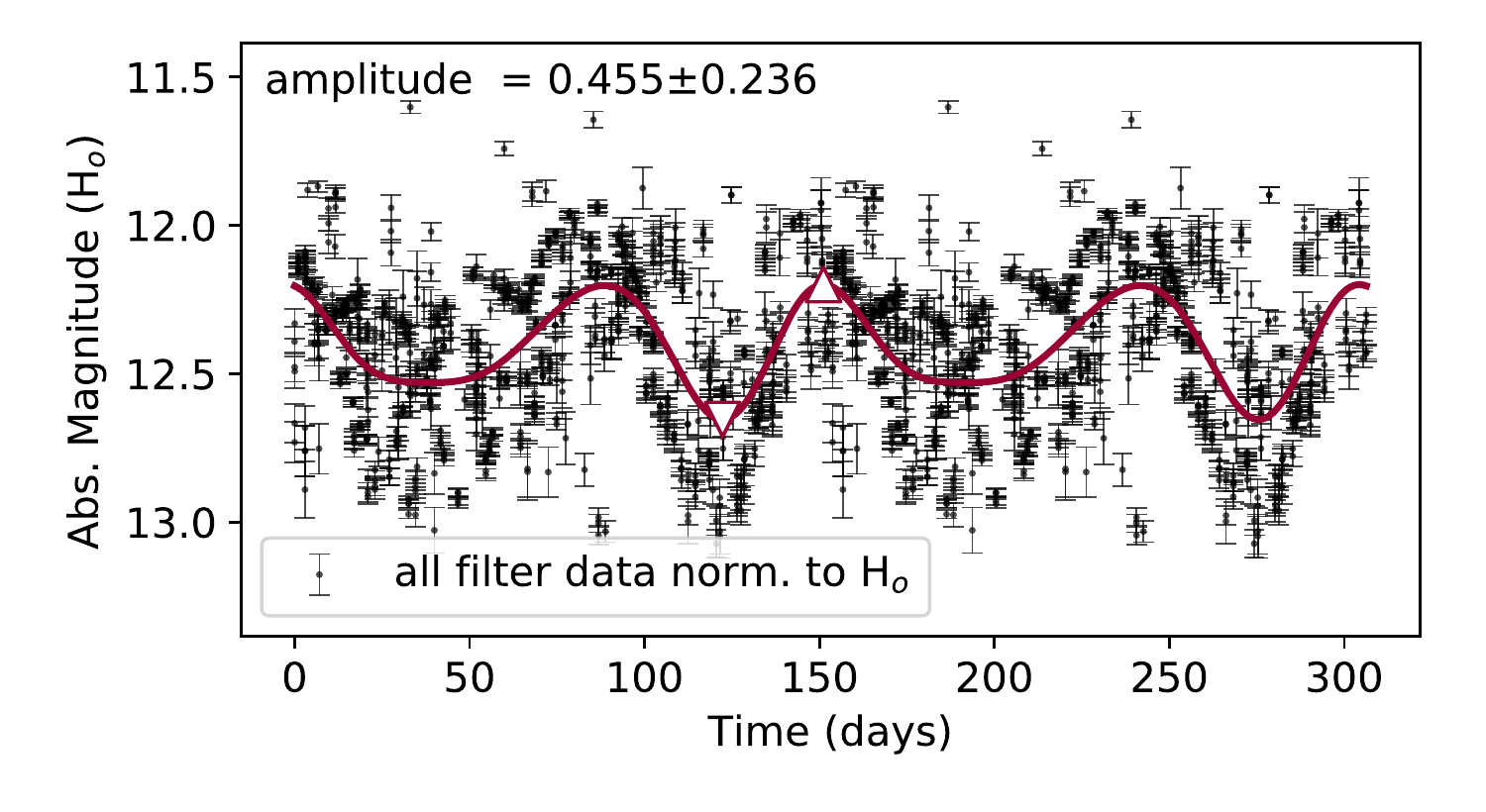}
    \caption{The combined ATLAS and ZTF filter data for asteroid (2831) Stevin folded with the period derived from the follow-up observation shown in Figure \ref{fig:Stevin_Druzhba_plot}.}
    \label{fig:Stevin_ATLAS_ZTF_alt}
\end{figure}

Our high-cadence follow-up observations are best fit for Druzhba (see Figure \ref{fig:Stevin_Druzhba_plot}) with a rotation period of 32~hours, confirming that the strongest 16-hour peak and not the slightly weaker (but still prominent) very long period peak is the 
best solution.

The follow-up observations for Stevin show an intra-night variation of around 0.1~mag, and a long time (90~days) trend that matches our ATLAS+ZTF rotation period of $>$3000~hours with a large amplitude around 1~magnitude (see Figure \ref{fig:Stevin_Druzhba_plot}). We thus confirm Stevin as a very slow rotator although the ATLAS/ZTF data produced a period solution of 3213~hours whereas the follow-up observations showed a period of 3717~hours. That said, 3717~hours corresponds to one of the adjacent peaks to our strongest periodogram peak (see middle panel of Figure \ref{fig:Stevin_ATLAS_ZTF} and Figure \ref{fig:Stevin_ATLAS_ZTF_alt} for the ATLAS/ZTF data folded at a period of 3717~hours).  

The follow-up PlaneWave data therefore provided further evidence that the relative peak strengths in our periodograms could be trusted, i.e., the strongest peak is the most likely  true period even if a similarly strong peak appears at a period that is orders of magnitude different.  The data also provided evidence that folding ATLAS/ZTF-type data at a long period (that turns out not to be the correct period but that also has a strong periodogram peak) can produce a clean and convincing folded light-curve so some caution should be taken when solely relying on clean folds of sparse data as evidence of a correct period. The counter argument also holds true as is the case for Stevin: large scatter in the folded light-curve is not an indication that the data is folded with the incorrect period. Effects like tumbling or light-curve amplitude changes across apparitions due to an
irregular shape
could also be the cause of the large 
scatter.    

\subsection{Simulations}

\begin{figure}
\includegraphics[width=\columnwidth]{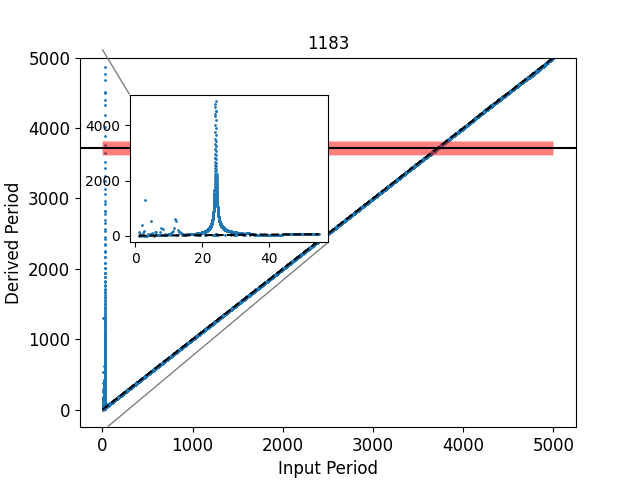}
    \caption{Period validation plot for asteroid (1183) 1930 DC. 
    Simulated input periods have a spacing of 0.25~hr except in the range 18--30~hr, where $\Delta P$ is $\SI{.01}{\hour}$ so more detail of the aliasing can be captured. The $\beta_P$ is adjusted to reflect the differing density of points by dividing the number of point in the range 18--30~hr by 25 (the ratio between the density elsewhere and the density in the range). A line with an error band is over-plotted at the period we derived for this object. The $\beta_P$ for this object is calculated from that band. \textit{Equivalent plots for all 39 objects will be available online.}}
\label{fig:1183_perval}
\end{figure}

The objects presented in this paper all have periodogram peaks related to the strong 1-day cadence of the ATLAS and ZTF data and not necessarily due to the rotation period of the asteroid. 
Here we present an additional method for assessing the validity of a periodogram solution. A period grid is created between a p$_{\text{min}}$ and p$_{\text{max}}$.
We simulate each asteroid in our candidate list by constructing a light curve for each period in the period grid using the same observing cadence as the 
observational record from ATLAS/ZTF.
At present, we use a simple sine curve; a future improvement would be to use a Fourier model, although a period would be needed to generate it and that might introduce a bias towards that period in the resulting plot. A periodogram is then generated for each step in the period grid and the strongest peak in the periodogram is taken as the derived period. 
Figure~\ref{fig:1183_perval} shows
our results for the derived periods in this simulation for asteroid 1183 Jutta (1930 DC).

\begin{figure}
    \centering
    \includegraphics[width=\columnwidth]{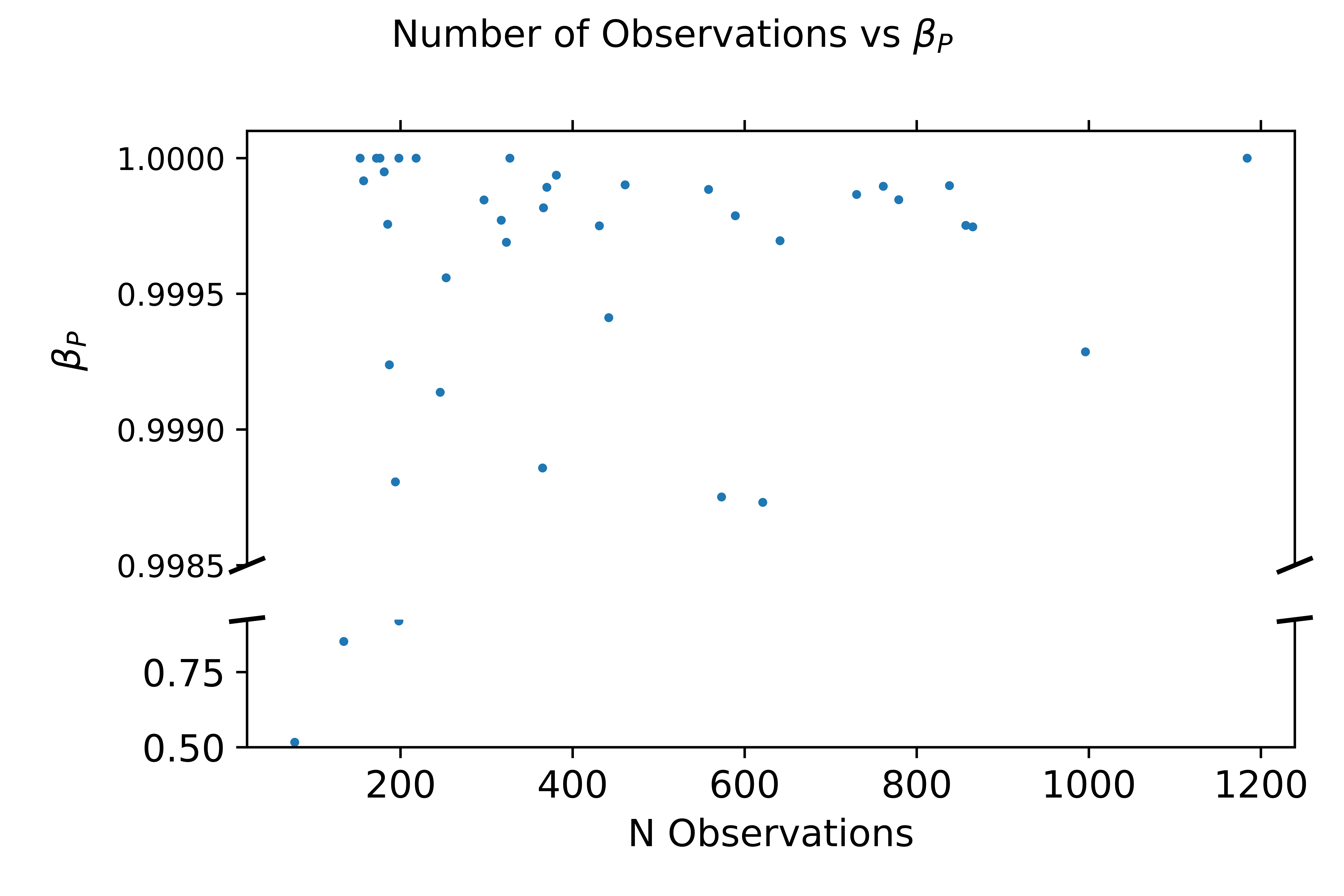}
    \caption{Shown are the number of observations an object has vs the $\beta_P$ with $P=$ the derived period for that object are included in Table~\ref{tab:data}. All objects have $\beta_P\gtrapprox0.99$ except
    311760 ($\beta_P\approx0.52$),
    116629 ($\beta_P\approx0.85$), 
    and 25629 ($\beta_P\approx0.92$).}
    \label{fig:perval_res}
\end{figure}

For a large amount of very good data with no time-based aliasing, we would expect that nearly all solutions lie along the line of slope~1, where the derived period is equal to the input simulated period. However, each of our real observed asteroids has imperfect data sets, with a limited number of observations, finite photometric errors, and the usual aliases associated with ground-based observing (i.e., 12~and 24~hours).
For each real asteroid (Table~\ref{tab:data}), we derive a period and uncertainty (black solid line and red horizontal band in Figure~\ref{fig:1183_perval}).
Within this band, we define the fraction of points on (or near) the line of slope~1 as $\beta_P$; larger values indicate solutions that we are most confident in.
For asteroid 1183, 
Figure~\ref{fig:1183_perval} shows
the result of this period validation experiment: for
our solution of $P=3713\pm\SI{101}{\hour}$ we find
$\beta_P = 0.999899$.
In other words, at the given derived period, our confidence that the derived period is the correct period is nearly 99.99\%.
We have carried out this same calculation for all the objects in Table~\ref{tab:data}, with the results shown in Figure~\ref{fig:perval_res}. It is important to note that for each object we use the actual ZTF/ATLAS observing cadence for that object, but with the full range of simulated input periods.
Out of our entire sample of 39 objects, 36~have $\beta_P\gtrapprox0.99$. 
Our confidence is especially high for objects with more than 200~observations.

\section{Discussion}
\label{sec:discussion}

The clear presence of many very slow rotators in the asteroid belt prompts the question of whether these rotation rates are primordial or the result of rotational evolution over the past 4.5~billion years. Both interpretations offer important constraints on the history of the asteroid belt.

\subsection{Non-principal axis rotation}

Very long rotation periods are often considered indicative of non-principal axis rotation. This is due to the low energy state of the system needing only slight disruption to have its rotation state significantly altered. In a non-monolithic body, tumbling results in a stress-strain cycle within the internal structure of the object. Excess energy is dissipated in this process and the object will undergo damping back to principal axis rotation. The timescale over which this effect occurs is the ``damping timescale'', $\tau$ (\citealt{pravec2002}).

\begin{figure}
	\includegraphics[width=\columnwidth]{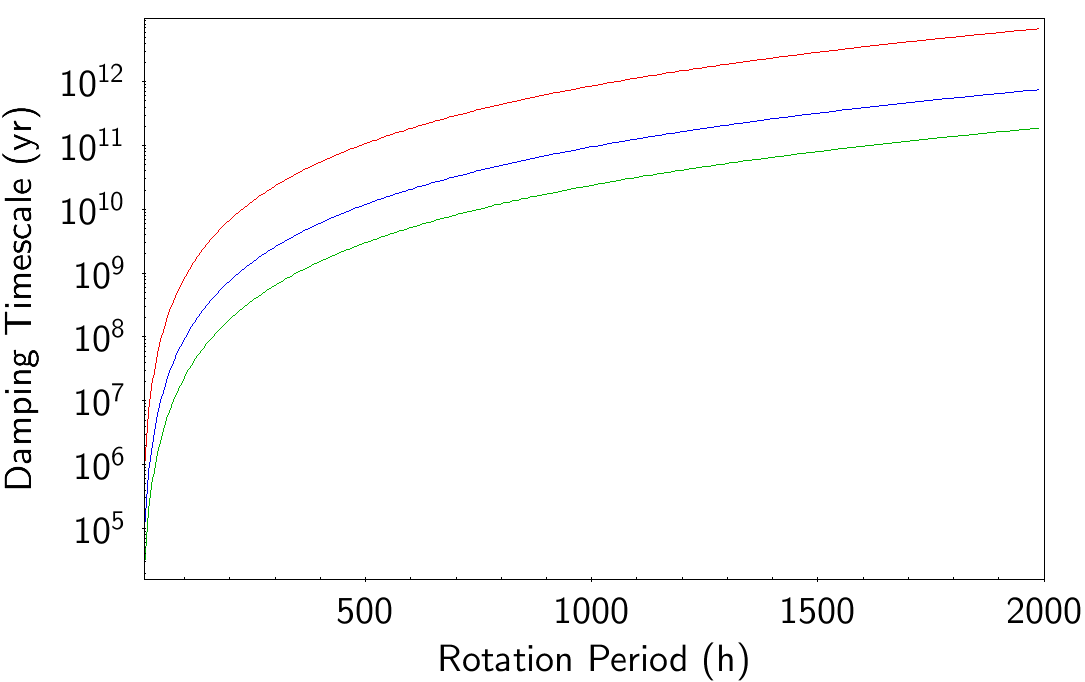}
    \caption{The damping timescale for a non-principal axis rotator to return to principal axis rotation for a given rotation period. The red, blue and green lines indicate the timescales for an object with diameter 5, 15 and 30 km respectively.}
    \label{fig:damping}
\end{figure}

As $\tau \propto P^{3}$ (where $P$ is the rotation period) it can be inferred that for a slow-rotator this timescale would be very long. Figure~\ref{fig:damping} shows the damping timescale for three tumblers of given diameter at a range of rotation rates. For even the largest case here of $D = 30$ km, for periods longer than $\sim$700 h the damping timescale is longer than the age of the Solar System; smaller and slower bodies have even longer timescales.
The targets presented in Table~\ref{tab:data} have 
periods longer than 1000~hours and
H~magnitudes in the range 12--16.5, implying sizes smaller than 30~km. Therefore, for all of these targets, the damping timescale is (much) more than 4.5~billion years. This suggests that 
if these objects were presently in non-principal axis rotation that this could have been their rotation state since their formation. 



Non-principal axis rotation will produce systematic changes in magnitude that can not be explained by the change in projected area about the object's rotation. In other words, there will be evidence of non-periodic structure in the folded light curves. Of the objects analysed here, only (2831) Stevin showed any evidence of this in the folded lightcurve. We therefore reject this explanation for the long periods we derive. 

Evidence of tumbling can be obscured by the sidereal-synodic period difference of the object. This is an effect causing shifting in rotation phase due to the object's motion, and will be more pronounced at slow rotations where objects move appreciably about their orbit within a single rotation. As we have a long baseline of observations spanning a wide range of viewing geometries, this is unlikely to be a significant effect.


\subsection{Thermal and collisional effects}

Another hypothesis for the presence of these ``slow rotators'' in the main belt population is that these are objects that have undergone significant spin-down by the Yarkovsky-O'Keefe-Radzievskii-Paddack (YORP) effect over their lifetime. The YORP effect acts to alter the rotation state of an asymmetric object due to thermal torques (\citealt{rubincam2000} ; \citealt{bottke2006}). These objects will then exist in a low-energy rotation state until they undergo a collision or are otherwise stimulated into non-principal axis rotation.

The rotation state of a kilometre-size asteroid with a typical rotation period $P = 6$ h can be altered by the YORP effect on timescales of order $10^{6}$ years. This timescale is proportional to the diameter of the body and inversely proportional to its rotation rate, implying short timescales for our $\sim$10~km bodies with rotation periods of thousands of hours.

\begin{figure}
	\includegraphics[width=\columnwidth]{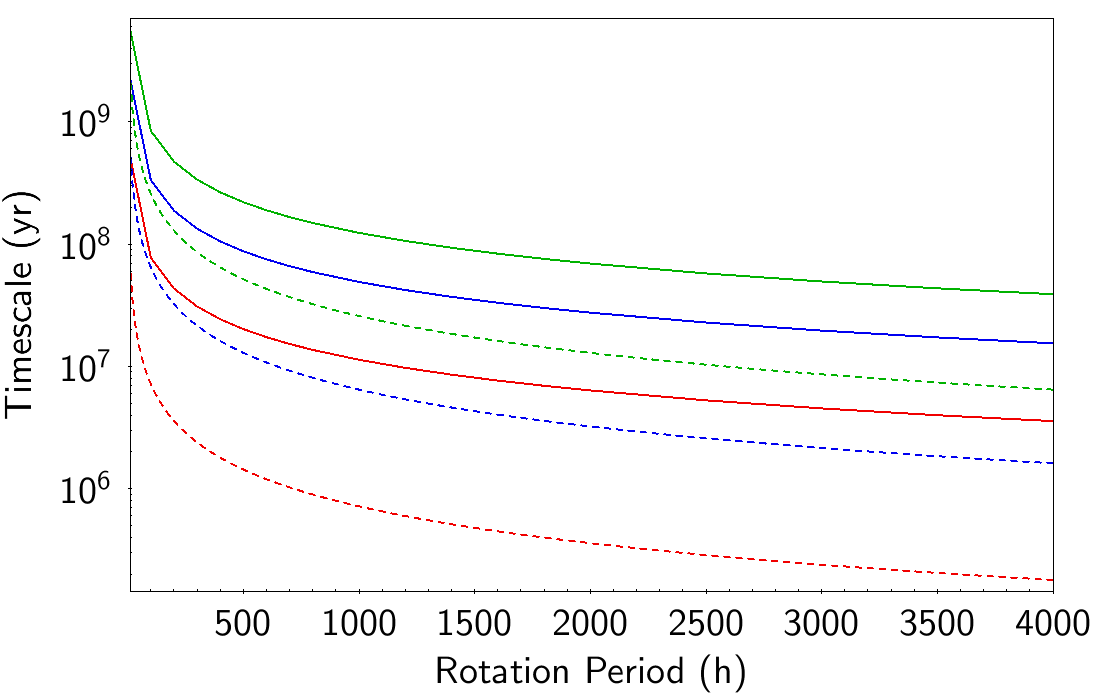}
    \caption{The collisional (solid) and YORP (dashed) timescales for an object of a given size at a range of rotation periods. The red, blue and green lines indicate the timescales for an object with diameter 5, 15 and 30 km respectively.}
    \label{fig:timescale_combo}
\end{figure}


Collisions can also affect the rotation rate of asteroids (either spinning up or down, depending on the sense of the collision with respect to the rotation pole).
For collisions to be the driving mechanism for these objects, this would also then require the collisional timescale to be shorter than that of YORP. From Figure~\ref{fig:timescale_combo} we see that for a given object the YORP timescale is always shorter than the collisional timescale with the disparity only growing with increasing rotation period. 

\subsection{Origin of very slow rotators}

These very slow rotators could in principle be in a non-principal axis rotation state because the timescale to damp back to principal axis rotation is very long. However, only one target shows possible evidence of non-principal axis rotation.
We therefore conclude that in general these very slow rotators are not in
non-principal axis rotation states.

The most plausible mechanism for the creation of these very slow rotators is therefore that these bodies had their rotations slowed by YORP spin-down.
In order to test this hypothesis,
we run a simple integration of several objects of given diameter and starting rotation period to test the timescales involved to reach slow rotation, i.e., the timescale is not constantly recomputed at each new period as the object's rotation evolves, it is instead only computed at the end of each YORP timescale ``step''. We consider objects with diameters 5, 15, and 30 km with initial rotation periods of 10, 100, and 1000 hours. Each object is assumed to be in a circular orbit at semi-major axis 2.5 au. We find that for the 5 km and 15 km object that each of the initial rotation periods is sufficient for the object to be driven to slow rotation states on timescales at least an order of magnitude shorter than the age of the Solar System. For a 30 km object representing the upper-end of our sample, we find that only an already slow rotation state ($P\gg 100$ h) can evolve to reach comparable rotation rates to our objects within this same time frame.

\begin{figure}
	\includegraphics[width=\columnwidth]{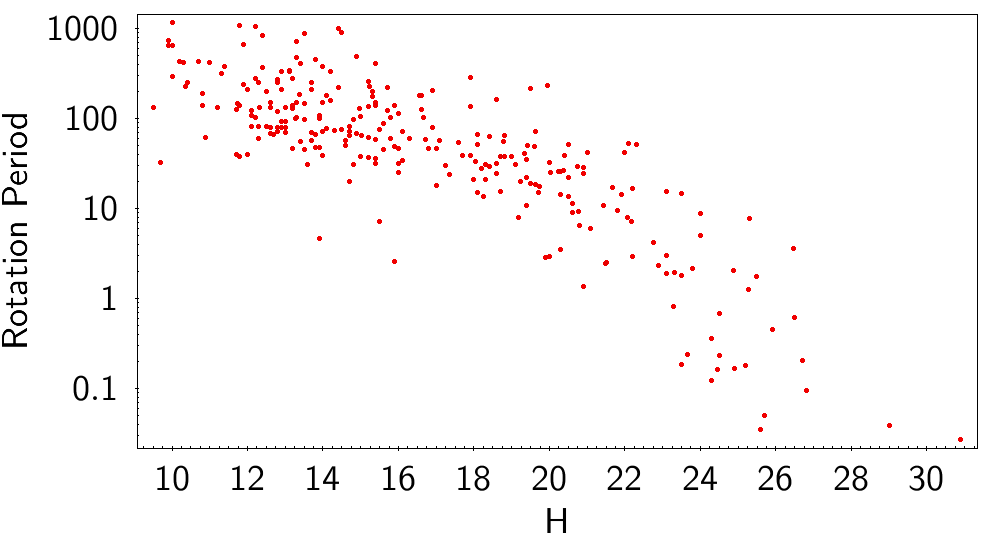}
    \caption{The rotation period of tumbling objects listed in the Light Curve Database plotted against their H magnitude.}
    \label{fig:lcdb_tumb}
\end{figure}

Collisions can effectively ``switch-off'' YORP: if an object's rotation is reset by a collision then YORP can not act on it to consistently increase or decrease its spin rate. The reverse is not true --- an object will still undergo a collision eventually regardless of YORP. Due to the low energy state of an object in a slow rotation state, a collision is likely to significantly alter the rotation state of the object, perhaps causing the onset of tumbling. This is in agreement with the overall relatively small fraction of slow rotators in the main belt, perhaps suggesting that this population of slow rotators is a ``transition region'' rather than an ``end-state''. This also appears to be in agreement with the correlation between rotation period and size for tumbling objects (Figure~\ref{fig:lcdb_tumb}; \citealt{Warner2009}), as larger objects will require a greater impact to evolve their rotation state and hence will be able to spin-down to slower rotation periods than their smaller counterparts. Similarly, large tumblers with short rotation periods will damp to principal-axis rotation on a short timescale.  

Thus, a reasonable set of initial conditions --- asteroids with unremarkable rotation periods ---
can produce a population of very slowly rotating asteroids as 
they are slowed by YORP over time.
Some asteroids (not identified here) that formerly had slow rotation periods have been reset or spun up by collisions.




\section{Conclusions}
\label{sec:conclusion}

We report here
the discovery of a new class of super-slow rotating asteroid, with periods of thousands of hours,
using observations from both ATLAS and ZTF. 
We present 39~such objects in this paper, and estimate that super slow rotating asteroids must be at least 0.4\% of the main belt asteroid population. These objects may have evolved to their current slow rotation rate through spin-down by the YORP effect with collisions eventually acting to reset this rotation state to non-principal axis rotation.
The forthcoming LSST will increase the number of known asteroids by a factor of five in a uniform and sparse survey that will be ideal for extending this work to more and much smaller asteroids.

\section*{Data availability}

The data underlying this article are available in the article and in its online supplementary material.

\section*{Acknowledgements}

This work has made use of data from the Asteroid Terrestrial-impact Last Alert System (ATLAS) project. ATLAS is primarily funded to search for near-Earth asteroids through NASA grants NN12AR55G, 80NSSC18K0284, and 80NSSC18K1575; byproducts of the NEA search include images and catalogs from the survey area.  The ATLAS science products have been made possible through the contributions of the University of Hawaii Institute for Astronomy, the Queen's University Belfast, the Space Telescope Science Institute, and the South African Astronomical Observatory.

This work is based in part on observations obtained with the Samuel Oschin 48-inch Telescope at the Palomar Observatory as part of the Zwicky Transient Facility project. ZTF is supported by the National Science Foundation under Grant No. AST-1440341 and a collaboration including Caltech, IPAC, the Weizmann Institute for Science, the Oskar Klein Center at Stockholm University, the University of Maryland, the University of Washington, Deutsches Elektronen-Synchrotron and Humboldt University, Los Alamos National Laboratories, the TANGO Consortium of Taiwan, the University of Wisconsin at Milwaukee, and Lawrence Berkeley National Laboratories. Operations are conducted by COO, IPAC, and UW.

This research made use of {\tt ccdproc}, an \texttt{Astropy} package for image reduction \citep{Craig2017}.
The {\tt PHOTOMETRYPIPELINE} package \citep{Mommert2017} was supported by NASA grants NNX15AE90G and NNX14AN82G and has been developed in the framework of the Mission Accessible Near-Earth Objects Survey (MANOS).

This work is partially supported by the South African National Research Foundation (NRF), the University of Cape Town, the Arizona Board of Regents' Regents Innovation Fund, the National Aeronautics and Space Administration (NASA), and by a grant from NASA's Office of the Chief Technologist.

This work made extensive use of \texttt{Python}, specifically the \texttt{NumPy} \citep{Harris2020}, \texttt{Astropy} \citep{Astropy2013,Astropy2018}, \texttt{Matplotlib} \citep{Hunter2007}, and \texttt{SciPy} \citep{2020SciPy-NMeth} packages.




\bibliographystyle{mnras}
\bibliography{citations} 

\begin{thebibliography}{}
\makeatletter
\relax
\def\mn@urlcharsother{\let\do\@makeother \do\$\do\&\do\#\do\^\do\_\do\%\do\~}
\def\mn@doi{\begingroup\mn@urlcharsother \@ifnextchar [ {\mn@doi@}
  {\mn@doi@[]}}
\def\mn@doi@[#1]#2{\def\@tempa{#1}\ifx\@tempa\@empty \href
  {http://dx.doi.org/#2} {doi:#2}\else \href {http://dx.doi.org/#2} {#1}\fi
  \endgroup}
\def\mn@eprint#1#2{\mn@eprint@#1:#2::\@nil}
\def\mn@eprint@arXiv#1{\href {http://arxiv.org/abs/#1} {{\tt arXiv:#1}}}
\def\mn@eprint@dblp#1{\href {http://dblp.uni-trier.de/rec/bibtex/#1.xml}
  {dblp:#1}}
\def\mn@eprint@#1:#2:#3:#4\@nil{\def\@tempa {#1}\def\@tempb {#2}\def\@tempc
  {#3}\ifx \@tempc \@empty \let \@tempc \@tempb \let \@tempb \@tempa \fi \ifx
  \@tempb \@empty \def\@tempb {arXiv}\fi \@ifundefined
  {mn@eprint@\@tempb}{\@tempb:\@tempc}{\expandafter \expandafter \csname
  mn@eprint@\@tempb\endcsname \expandafter{\@tempc}}}

\bibitem[\protect\citeauthoryear{{Astropy Collaboration} et~al.,}{{Astropy
  Collaboration} et~al.}{2013}]{Astropy2013}
{Astropy Collaboration} et~al., 2013, \mn@doi [\aap]
  {10.1051/0004-6361/201322068}, \href
  {https://ui.adsabs.harvard.edu/abs/2013A&A...558A..33A} {558, A33}

\bibitem[\protect\citeauthoryear{{Astropy Collaboration} et~al.,}{{Astropy
  Collaboration} et~al.}{2018}]{Astropy2018}
{Astropy Collaboration} et~al., 2018, \mn@doi [\aj] {10.3847/1538-3881/aabc4f},
  \href {https://ui.adsabs.harvard.edu/abs/2018AJ....156..123A} {156, 123}

\bibitem[\protect\citeauthoryear{{Bellm} \& {Kulkarni}}{{Bellm} \&
  {Kulkarni}}{2017}]{Bellm2017}
{Bellm} E.,  {Kulkarni} S.,  2017, \mn@doi [Nature Astronomy]
  {10.1038/s41550-017-0071}, \href
  {https://ui.adsabs.harvard.edu/abs/2017NatAs...1E..71B} {1, 0071}

\bibitem[\protect\citeauthoryear{{Bolin} et~al.,}{{Bolin}
  et~al.}{2020}]{Bolin2021}
{Bolin} B.~T.,  et~al., 2020, arXiv e-prints, \href
  {https://ui.adsabs.harvard.edu/abs/2020arXiv201103782B} {p. arXiv:2011.03782}

\bibitem[\protect\citeauthoryear{{Bottke}, {Vokrouhlick{\'y}}, {Rubincam}  \&
  {Nesvorn{\'y}}}{{Bottke} et~al.}{2006}]{bottke2006}
{Bottke} William~F. J.,  {Vokrouhlick{\'y}} D.,  {Rubincam} D.~P.,
  {Nesvorn{\'y}} D.,  2006, \mn@doi [Annual Review of Earth and Planetary
  Sciences] {10.1146/annurev.earth.34.031405.125154}, \href
  {https://ui.adsabs.harvard.edu/abs/2006AREPS..34..157B} {34, 157}

\bibitem[\protect\citeauthoryear{{Bowell}, {Hapke}, {Domingue}, {Lumme},
  {Peltoniemi}  \& {Harris}}{{Bowell} et~al.}{1989}]{Bowell1989}
{Bowell} E.,  {Hapke} B.,  {Domingue} D.,  {Lumme} K.,  {Peltoniemi} J.,
  {Harris} A.~W.,  1989, in {Binzel} R.~P.,  {Gehrels} T.,   {Matthews} M.~S.,
  eds, Asteroids II. pp 524--556

\bibitem[\protect\citeauthoryear{Craig et~al.,}{Craig et~al.}{2017}]{Craig2017}
Craig M.,  et~al., 2017, astropy/ccdproc: v1.3.0.post1,
  \mn@doi{10.5281/zenodo.1069648}, \url
  {https://doi.org/10.5281/zenodo.1069648}

\bibitem[\protect\citeauthoryear{Duev et~al.,}{Duev et~al.}{2019a}]{Duev2019}
Duev D.~A.,  et~al., 2019a, \mn@doi [Monthly Notices of the Royal Astronomical
  Society] {10.1093/mnras/stz1096}, 486, 4158

\bibitem[\protect\citeauthoryear{Duev et~al.,}{Duev et~al.}{2019b}]{Duev2019b}
Duev D.~A.,  et~al., 2019b, \mn@doi [Monthly Notices of the Royal Astronomical
  Society] {10.1093/mnras/stz2357}, 489, 3582–3590

\bibitem[\protect\citeauthoryear{{Dymock}}{{Dymock}}{2007}]{Dymock2007}
{Dymock} R.,  2007, Journal of the British Astronomical Association, \href
  {https://ui.adsabs.harvard.edu/abs/2007JBAA..117..342D} {117, 342}

\bibitem[\protect\citeauthoryear{{Erasmus}, {Mommert}, {Trilling},
  {Sickafoose}, {van Gend}  \& {Hora}}{{Erasmus} et~al.}{2017}]{Erasmus2017}
{Erasmus} N.,  {Mommert} M.,  {Trilling} D.~E.,  {Sickafoose} A.~A.,  {van
  Gend} C.,   {Hora} J.~L.,  2017, \mn@doi [\aj] {10.3847/1538-3881/aa88be},
  \href {https://ui.adsabs.harvard.edu/abs/2017AJ....154..162E} {154, 162}

\bibitem[\protect\citeauthoryear{{Erasmus} et~al.,}{{Erasmus}
  et~al.}{2020}]{Erasmus2020}
{Erasmus} N.,  et~al., 2020, \mn@doi [\apjs] {10.3847/1538-4365/ab5e88}, \href
  {https://ui.adsabs.harvard.edu/abs/2020ApJS..247...13E} {247, 13}

\bibitem[\protect\citeauthoryear{{Harris} et~al.,}{{Harris}
  et~al.}{2020}]{Harris2020}
{Harris} C.~R.,  et~al., 2020, \mn@doi [\nat] {10.1038/s41586-020-2649-2},
  \href {https://ui.adsabs.harvard.edu/abs/2020Natur.585..357H} {585, 357}

\bibitem[\protect\citeauthoryear{{Heinze} et~al.,}{{Heinze}
  et~al.}{2018}]{Heinze2018}
{Heinze} A.~N.,  et~al., 2018, \mn@doi [\aj] {10.3847/1538-3881/aae47f}, \href
  {https://ui.adsabs.harvard.edu/abs/2018AJ....156..241H} {156, 241}

\bibitem[\protect\citeauthoryear{Heinze et~al.,}{Heinze
  et~al.}{2021}]{Heinze2021}
Heinze A.~N.,  et~al., 2021, \mn@doi [The Planetary Science Journal]
  {10.3847/psj/abd325}, 2, 12

\bibitem[\protect\citeauthoryear{{Hunter}}{{Hunter}}{2007}]{Hunter2007}
{Hunter} J.~D.,  2007, \mn@doi [Computing in Science Engineering]
  {10.1109/MCSE.2007.55}, 9, 90

\bibitem[\protect\citeauthoryear{{Ip} et~al.,}{{Ip} et~al.}{2020}]{Ip2020}
{Ip} W.~H.,  et~al., 2020, arXiv e-prints, \href
  {https://ui.adsabs.harvard.edu/abs/2020arXiv200904125I} {p. arXiv:2009.04125}

\bibitem[\protect\citeauthoryear{Ivezi{\'{c}} et~al.,}{Ivezi{\'{c}}
  et~al.}{2001}]{Ivezic2001}
Ivezi{\'{c}} {\v{Z}}.,  et~al., 2001, \mn@doi [The Astronomical Journal]
  {10.1086/323452}, 122, 2749

\bibitem[\protect\citeauthoryear{{Lomb}}{{Lomb}}{1976}]{Lomb1976}
{Lomb} N.~R.,  1976, \mn@doi [\apss] {10.1007/BF00648343}, \href
  {https://ui.adsabs.harvard.edu/abs/1976Ap&SS..39..447L} {39, 447}

\bibitem[\protect\citeauthoryear{{Mahlke}, {Carry}  \& {Denneau}}{{Mahlke}
  et~al.}{2021}]{Mahlke2021}
{Mahlke} M.,  {Carry} B.,   {Denneau} L.,  2021, \mn@doi [\icarus]
  {10.1016/j.icarus.2020.114094}, \href
  {https://ui.adsabs.harvard.edu/abs/2021Icar..35414094M} {354, 114094}

\bibitem[\protect\citeauthoryear{{Masci} et~al.,}{{Masci}
  et~al.}{2019}]{Masci2019}
{Masci} F.~J.,  et~al., 2019, \mn@doi [\pasp] {10.1088/1538-3873/aae8ac}, \href
  {https://ui.adsabs.harvard.edu/abs/2019PASP..131a8003M} {131, 018003}

\bibitem[\protect\citeauthoryear{{Matheson} et~al.,}{{Matheson}
  et~al.}{2020}]{Matheson2020}
{Matheson} T.,  et~al., 2020, arXiv e-prints, \href
  {https://ui.adsabs.harvard.edu/abs/2020arXiv201112385M} {p. arXiv:2011.12385}

\bibitem[\protect\citeauthoryear{{McNeill} et~al.,}{{McNeill}
  et~al.}{2016}]{McNeill2016}
{McNeill} A.,  et~al., 2016, \mn@doi [\mnras] {10.1093/mnras/stw847}, \href
  {https://ui.adsabs.harvard.edu/abs/2016MNRAS.459.2964M} {459, 2964}

\bibitem[\protect\citeauthoryear{{McNeill}, {Hora}, {Gustafsson}, {Trilling}
  \& {Mommert}}{{McNeill} et~al.}{2019}]{McNeill2019}
{McNeill} A.,  {Hora} J.~L.,  {Gustafsson} A.,  {Trilling} D.~E.,   {Mommert}
  M.,  2019, \mn@doi [\aj] {10.3847/1538-3881/ab0e6e}, \href
  {https://ui.adsabs.harvard.edu/abs/2019AJ....157..164M} {157, 164}

\bibitem[\protect\citeauthoryear{McNeill et~al.,}{McNeill
  et~al.}{2021}]{McNeill2021}
McNeill A.,  et~al., 2021, \mn@doi [The Planetary Science Journal]
  {10.3847/psj/abcccd}, 2, 6

\bibitem[\protect\citeauthoryear{Moln{\'{a}}r et~al.,}{Moln{\'{a}}r
  et~al.}{2018}]{Molnar2018}
Moln{\'{a}}r L.,  et~al., 2018, \mn@doi [The Astrophysical Journal Supplement
  Series] {10.3847/1538-4365/aaa1a1}, 234, 37

\bibitem[\protect\citeauthoryear{{Mommert}}{{Mommert}}{2017}]{Mommert2017}
{Mommert} M.,  2017, \mn@doi [Astronomy and Computing]
  {10.1016/j.ascom.2016.11.002}, \href
  {https://ui.adsabs.harvard.edu/abs/2017A&C....18...47M} {18, 47}

\bibitem[\protect\citeauthoryear{{Mommert} et~al.,}{{Mommert}
  et~al.}{2019}]{Mommert_2019}
{Mommert} M.,  et~al., 2019, \mn@doi [The Journal of Open Source Software]
  {10.21105/joss.01426}, \href
  {https://ui.adsabs.harvard.edu/abs/2019JOSS....4.1426M} {4, 1426}

\bibitem[\protect\citeauthoryear{P{\'{a}}l et~al.,}{P{\'{a}}l
  et~al.}{2020}]{Polnar2020}
P{\'{a}}l A.,  et~al., 2020, \mn@doi [The Astrophysical Journal Supplement
  Series] {10.3847/1538-4365/ab64f0}, 247, 26

\bibitem[\protect\citeauthoryear{{Patterson} et~al.,}{{Patterson}
  et~al.}{2019}]{Patterson2019}
{Patterson} M.~T.,  et~al., 2019, \mn@doi [\pasp] {10.1088/1538-3873/aae904},
  \href {https://ui.adsabs.harvard.edu/abs/2019PASP..131a8001P} {131, 018001}

\bibitem[\protect\citeauthoryear{{Polakis}}{{Polakis}}{2019}]{Polakis2019}
{Polakis} T.,  2019, Minor Planet Bulletin, \href
  {https://ui.adsabs.harvard.edu/abs/2019MPBu...46..132P} {46, 132}

\bibitem[\protect\citeauthoryear{{Pravec}, {Harris}  \& {Michalowski}}{{Pravec}
  et~al.}{2002}]{pravec2002}
{Pravec} P.,  {Harris} A.~W.,   {Michalowski} T.,  2002, {Asteroid Rotations}.
pp 113--122

\bibitem[\protect\citeauthoryear{{Rubincam}}{{Rubincam}}{2000}]{rubincam2000}
{Rubincam} D.~P.,  2000, \mn@doi [\icarus] {10.1006/icar.2000.6485}, \href
  {https://ui.adsabs.harvard.edu/abs/2000Icar..148....2R} {148, 2}

\bibitem[\protect\citeauthoryear{{Scargle}}{{Scargle}}{1982}]{Scargle1982}
{Scargle} J.~D.,  1982, \mn@doi [\apj] {10.1086/160554}, \href
  {https://ui.adsabs.harvard.edu/abs/1982ApJ...263..835S} {263, 835}

\bibitem[\protect\citeauthoryear{{Schemel} \& {Brown}}{{Schemel} \&
  {Brown}}{2020}]{Schemel2021}
{Schemel} M.,  {Brown} M.~E.,  2020, arXiv e-prints, \href
  {https://ui.adsabs.harvard.edu/abs/2020arXiv201101329S} {p. arXiv:2011.01329}

\bibitem[\protect\citeauthoryear{{Smith} et~al.,}{{Smith}
  et~al.}{2014}]{Smith2014}
{Smith} R.~M.,  et~al., 2014, in {Ramsay} S.~K.,  {McLean} I.~S.,   {Takami}
  H.,  eds,  Society of Photo-Optical Instrumentation Engineers (SPIE)
  Conference Series Vol. 9147, Ground-based and Airborne Instrumentation for
  Astronomy V. p. 914779, \mn@doi{10.1117/12.2070014}

\bibitem[\protect\citeauthoryear{{Smith} et~al.,}{{Smith}
  et~al.}{2020}]{Smith2020}
{Smith} K.~W.,  et~al., 2020, arXiv e-prints, \href
  {https://ui.adsabs.harvard.edu/abs/2020arXiv200309052S} {p. arXiv:2003.09052}

\bibitem[\protect\citeauthoryear{{Stephens}}{{Stephens}}{2011}]{Stephens2011a}
{Stephens} R.~D.,  2011, Minor Planet Bulletin, \href
  {https://ui.adsabs.harvard.edu/abs/2011MPBu...38..211S} {38, 211}

\bibitem[\protect\citeauthoryear{{Tonry} et~al.,}{{Tonry}
  et~al.}{2018a}]{Tonry2018a}
{Tonry} J.~L.,  et~al., 2018a, \mn@doi [\pasp] {10.1088/1538-3873/aabadf},
  \href {https://ui.adsabs.harvard.edu/abs/2018PASP..130f4505T} {130, 064505}

\bibitem[\protect\citeauthoryear{{Tonry} et~al.,}{{Tonry}
  et~al.}{2018b}]{Tonry2018b}
{Tonry} J.~L.,  et~al., 2018b, \mn@doi [\apj] {10.3847/1538-4357/aae386}, \href
  {https://ui.adsabs.harvard.edu/abs/2018ApJ...867..105T} {867, 105}

\bibitem[\protect\citeauthoryear{Virtanen et~al.,}{Virtanen
  et~al.}{2020a}]{SciPy2020}
Virtanen P.,  et~al., 2020a, \mn@doi [Nature Methods]
  {10.1038/s41592-019-0686-2}, \href {https://rdcu.be/b08Wh} {17, 261}

\bibitem[\protect\citeauthoryear{Virtanen et~al.,}{Virtanen
  et~al.}{2020b}]{2020SciPy-NMeth}
Virtanen P.,  et~al., 2020b, \mn@doi [Nature Methods]
  {10.1038/s41592-019-0686-2}, \href {https://rdcu.be/b08Wh} {17, 261}

\bibitem[\protect\citeauthoryear{{Warner}, {Harris}  \& {Pravec}}{{Warner}
  et~al.}{2009}]{Warner2009}
{Warner} B.~D.,  {Harris} A.~W.,   {Pravec} P.,  2009, \mn@doi [\icarus]
  {10.1016/j.icarus.2009.02.003}, \href
  {https://ui.adsabs.harvard.edu/abs/2009Icar..202..134W} {202, 134}

\bibitem[\protect\citeauthoryear{Ye et~al.,}{Ye et~al.}{2019a}]{Ye2019a}
Ye Q.,  et~al., 2019a, \mn@doi [Publications of the Astronomical Society of the
  Pacific] {10.1088/1538-3873/ab1b18}, 131, 078002

\bibitem[\protect\citeauthoryear{Ye et~al.,}{Ye et~al.}{2019b}]{Ye2019b}
Ye Q.,  et~al., 2019b, \mn@doi [The Astrophysical Journal]
  {10.3847/2041-8213/ab0f3c}, 874, L16

\bibitem[\protect\citeauthoryear{{{\v{D}}urech}, {Tonry}, {Erasmus}, {Denneau},
  {Heinze}, {Flewelling}  \& {Van{\v{c}}o}}{{{\v{D}}urech}
  et~al.}{2020}]{Durech2020}
{{\v{D}}urech} J.,  {Tonry} J.,  {Erasmus} N.,  {Denneau} L.,  {Heinze} A.~N.,
  {Flewelling} H.,   {Van{\v{c}}o} R.,  2020, \mn@doi [\aap]
  {10.1051/0004-6361/202037729}, \href
  {https://ui.adsabs.harvard.edu/abs/2020A&A...643A..59D} {643, A59}

\makeatother
\end{thebibliography}



\bsp	
\label{lastpage}
\end{document}